\documentclass[pdflatex,sn-mathphys-ay]{sn-jnl}
\usepackage{siunitx}
\usepackage{graphicx}%
\usepackage{multirow}%
\usepackage{amsmath,amssymb,amsfonts}%
\usepackage{amsthm}%
\usepackage{mathrsfs}%
\usepackage[title]{appendix}%
\usepackage{xcolor}%
\usepackage{textcomp}%
\usepackage{manyfoot}%
\usepackage{booktabs}%
\usepackage{algorithm}%
\usepackage{algorithmicx}%
\usepackage{algpseudocode}%
\usepackage{listings}%
\usepackage{braket}
\usepackage{lipsum}
\usepackage{amsmath,amssymb,amsfonts}
\usepackage{algorithm}
\usepackage{algpseudocode}
\usepackage{graphicx}
\usepackage{textcomp}
\usepackage{xcolor}
\usepackage{braket}
\usepackage{enumerate}
\usepackage{soul}
\theoremstyle{thmstyleone}%

\theoremstyle{thmstyletwo}%
\usepackage[x11names]{xcolor}
\usepackage[color]{changebar}
\usepackage{xparse}
\usepackage{ifthen}
\usepackage{marginnote}

\usepackage[most]{tcolorbox}
\usepackage{xcolor}

\theoremstyle{thmstylethree}%

\begin{document}

\title[Hybrid Reward-Driven Reinforcement Learning for Efficient Quantum Circuit Synthesis]{Hybrid Reward-Driven Reinforcement Learning for Efficient Quantum Circuit Synthesis}

\author*[1]{\fnm{Sara} \sur{Giordano}}\email{sgiordan@ucm.es}

\author[1]{\fnm{Kornikar} \sur{Sen}}\email{skornikar@ucm.es}

\author[1,2]{\fnm{Miguel A.} \sur{Martin-Delgado}}\email{mardel@fis.ucm.es}

\affil*[1]{\orgdiv{Departamento de Física Teórica, }, \orgname{Universidad Complutense de Madrid}, \orgaddress{\street{Plaza de las Ciencias 1. Ciudad Universitaria}, \city{Madrid}, \postcode{28040}, \state{Madrid}, \country{Spain}}}

\affil[2]{\orgdiv{CCS-Center for Computational Simulation}, \orgname{Campus de Montegancedo Universidad Politecnica de Madrid (UPM)}, \orgaddress{\street{Av. de Montepríncipe,}, \city{Boadilla del Monte}, \postcode{28660}, \state{Madrid}, \country{Spain}}}

\abstract{A reinforcement learning (RL) framework is introduced for the efficient synthesis of quantum circuits that generate specified target quantum states from a fixed initial state, addressing a central challenge in both the Noisy Intermediate-Scale Quantum (NISQ) era and future fault-tolerant quantum computing. The approach utilizes tabular Q-learning, based on action sequences, within a discretized quantum state space, to effectively manage the exponential growth of the space dimension. The framework introduces a hybrid reward mechanism, combining a static, domain-informed reward that guides the agent toward the target state with customizable dynamic penalties that discourage inefficient circuit structures such as gate congestion and redundant state revisits. This is a circuit-aware reward, in contrast to the current trend of works on this topic, which are primarily fidelity-based. By leveraging sparse matrix representations and state-space discretization, the method enables practical navigation of high-dimensional environments while minimizing computational overhead. Benchmarking on graph-state preparation tasks for up to seven qubits, we demonstrate that the algorithm consistently discovers minimal-depth circuits with optimized gate counts. Moreover, extending the framework to a universal gate set still yields low depth circuits, highlighting the algorithm’s robustness and adaptability. The results confirm that this RL-driven approach, with our completely circuit-aware method, efficiently explores the complex quantum state space and synthesizes near-optimal quantum circuits, providing a resource-efficient foundation for quantum circuit optimization.}


\maketitle

\section{Introduction}\label{sec1}
\label{sec:1}
The design of optimized quantum circuits is a crucial task in the current NISQ era and for the future of fault-tolerant quantum computing. Various methods based on combinatorial optimization and mathematically rigorous techniques have been employed for the interest in quantum computing over the past decades. Alongside these standard approaches, the use of Artificial Intelligence (AI)--in particular, Reinforcement Learning (RL) techniques (\cite{RusselAI,sutton2018reinforcement, wiering2012reinforcement})--has gained importance in the field, contributing to the generation of new circuits with reduced depth, fewer gates, or improved geometric structures tailored to current hardware constraints (\cite{RL4qubits, moflic2023}).

RL has been shown to discover high-quality quantum control protocols (\cite{Bukov2018}) and, more broadly, machine learning has become a key tool in the quantum domain (\cite{Dunjko_2018}), building on the success of deep RL (DRL) in complex decision-making (\cite{Mnih2015}). Within quantum information, deep RL helps in optimizing the techniques for quantum error correction (\cite{Nautrup2019optimizingquantum, Andreasson2019quantumerror}) and even automated experimental design (\cite{Melkinov2018}). DRL has also enabled universal quantum control and efficient exploration of control landscapes (\cite{Cramer2010EfficientQST, Niu2019UniversalControl, Dalgaard2020AlphaZero}), while recent works have addressed many-body control with tensor-network–assisted RL (\cite{Metz2023SelfCorrecting}), supervised state preparation (\cite{Selig2023}), and DRL-based compilation and synthesis (\cite{Chen_2024, Rietsch2024, Yu_knowledge_driven_2025, SUN2025}).

In this work, we address the problem of synthesizing quantum circuits that generate a target quantum state from a fixed initial state by means of reinforcement learning. Specifically, we propose a tabular Q-learning framework (\cite{watkins1992q}) in which the agent learns action sequences over a discretized representation of quantum states. In contrast to most deep-RL state-preparation algorithms---where the reward signal is primarily fidelity-based---we define a reward linked directly to the circuit construction. The learning process is guided by a hybrid reward mechanism: we use a sparse static reward precomputed offline, guiding the agent to the target state, and dynamic penalties calculated during the learning process. These penalties encode other circuit constraints aiming to optimize simultaneously more features of the final circuit, such as minimal depth or minimizing costly gates (\cite{ng1999policy}). This sign separation ensures that penalties will never define a competing objective: they only refine the path leading to the target state. The same reward that drives preparation simultaneously promotes compact circuits—an alignment we validate in the few-qubit regime.

Due to the requirement of labeling each states the algorithm can visit, we restrict ourselves to the set of States With Equal-amplitude and Encoded-phase Terms (SWEET), described in Sec.~\ref{subsec:SWEET}. Various resourceful states, such as the graph states (\cite{SChlingemannGraphStates, RaussendorfGraphStates}), GHZ states (\cite{Kafatos:1989bu}), W states (\cite{PhysRevA.62.062314}), fall into this set. In Sec.~\ref{sub:benchmarking_intro}, we construct the optimal circuits for preparation of the graph states, which are precisely represented by the SWEET states we define. We benchmark our algorithm by showing the circuits for preparation of graph states have the minimum depth, supported by Vizing's theorem (\cite{MISRA1992131}).

While remaining in the SWEET states category, we can handle treating different phases in the state superposition. The phases are parameterized, introducing a particular integer-type variable, which can be referred to as the phase variable. Limitations appear when we aim to tune different amplitudes for the desired states, since different amplitudes are explicitly excluded by the SWEET states category. This obstacle could also be overcome by tuning specific gates, within the overall optimized circuit, in post-production, as it was proposed in our previous work \cite{RL4qubits}. To consider states with unequal amplitudes within the algorithm, new amplitude variables can be introduced in addition to the phase variable. However, introduction of the new variables will increase the size of the set of states, which is already considerable. We remark that, despite the problem of the curse of dimensionality remains, we manage to design our original reward structure maintaining the problem tractable with our discretization of the state space and through designing an effective exploration.

In Sec.~\ref{subsec:general_result_intro}, we moved to the creation of optimized circuits with a set of universal gates, always targeting SWEET states. While for the graph states we obtained circuits preparing a state which corresponds exactly to the target one in the SWEET states, with the universal gates set the obtained circuits do not always produce the targeted state, thus we compute the fidelity between this latter and the output state of the circuit. In the reported example in the results in Section \ref{subsec:optimized_circuits_universal_set} we successfully prepare a state with $0.97$ fidelity with respect to the targeted one. This mismatch constitutes a limitation to our capability of tuning the states amplitudes in the universal gates set case, but it does not interfere with the optimality of the circuit construction.

Although the quantum state space grows doubly exponentially with the number of qubits and the action space scales polynomially, our use of discretization and sparse matrix representations (\cite{li2009sparse}) keeps the problem tractable under certain assumptions. A Q-learning setup, with a fixed initial state and finite action sequences ensures efficient exploration despite the vastness of the state space (\cite{dulac2019challenges}). Moreover, we introduce multiple strata of static reward around the target state, in order to reduce its sparsity, and thereby improving the algorithm's performances. To limit offline computation, we restrict our benchmarks to systems of up to $7$ qubits.

The paper is organized as follows. In Sec.~\ref{sec:2}, we review the relevant background on quantum state representations, gate sets, and circuit metrics. Sec.~\ref{sec:3} presents the Q-learning algorithm in detail, including the reward design and exploration strategy. Sec.~\ref{sec:4} outlines the implementation aspects while Subsection~\ref{sub:benchmarking_intro} focuses on graph-state preparation tasks, and Subsection~\ref{subsec:optimized_circuits_universal_set} extends the framework to more general target states. Finally, conclusions and future directions are discussed in Sec.~\ref{sec:5}.

\section{Preliminaries}\label{sec:2}
Before going into the technical details of the algorithm used to build efficient circuits, let us first discuss a few basic concepts which will be used in the remaining part of the paper.

\subsection{States With Equal-amplitude and Encoded-phase Terms}\label{subsec:SWEET}
To utilize the RL method for constructing an appropriate circuit that is capable of producing the exact form of an $n$-qubit target state, we should be able to label all possible $n$-qubit states contained in the $n$-qubit Hilbert, $\mathcal{H}_n$. Since $\mathcal{H}_n$ consists of an infinite number of states, it is infeasible to label each of them. Hence, instead of considering the entire Hilbert space, $\mathcal{H}_n$, we focus on a specific finite subset $\mathcal{S}_n$ of $\mathcal{H}_n$, which includes all the states $\ket{\psi_n}$ that can be expressed as
\begin{equation}\label{eq_SWEET}
    \ket{\psi_n}=\frac{1}{\sqrt{N'}}\sum_{j=1}^{N} \alpha_j\ket{x_j},
\end{equation}
where $\ket{x_j}$ is an element of the computational basis, $\mathcal{B}_n\equiv \{\ket{0},\ket{1}\}^{\otimes n}$ of $\mathcal{H}_n$ $\forall$ $j$, $\alpha_j\in\mathcal{P}_M:=\{e^{i2m\pi/M}\}_{m\in \mathbb{N}}$ is a complex phase, $N\in \mathbb{N}$ specifies the number of terms in superposition and $N'$ is the normalization constant. Here, $M\in\mathbb{N}$ characterizes the size of the finite set, $\mathcal{P}_M$, and correspondingly the discreteness of the complex phases. 
Care must be taken to recognize that the index $j$ labels the different terms in the expansion of $\ket{\psi_n}$. Importantly, the equality $\ket{x_j}= \ket{x_{j'}}$ does not necessarily imply that $j= j'$. Consequently, the total number of unique indices $N'$ is, in general, not equal to $N$.

This construction allows us to represent states with equal amplitude terms and discrete phases, deliberately neglecting the full complex amplitudes. We refer to such states as States With Equal-amplitude and Encoded-phase Terms (SWEET). We always choose the discreteness, $M$, of the phase set, $\mathcal{P}_M$, in such a way that the cardinality, $P_M$, of $\mathcal{P}_M$ can be written as $P_M=M=2^p$ where $p\in\mathbb{N}$. Although this limits the expressive power of $\mathcal{S}_n$, it is particularly well suited for representing certain classes of quantum states, such as graph states (\cite{SChlingemannGraphStates, RaussendorfGraphStates}), GHZ states (\cite{Kafatos:1989bu}), W states (\cite{PhysRevA.62.062314}), and other highly entangled states with uniform amplitudes, where the relative phase carries all relevant information. 

Most importantly, this simplification enables a compact representation of quantum states and allows for an efficient integration with our reinforcement learning-based circuit synthesis framework. It leads to a finite and discrete state space suitable for encoding in Q-learning. Moreover, it supports the application of a certain universal gate set (e.g., ${\text{CNOT}, H, T}$) in an efficient and tractable manner. Thus, in this work, we limit our study to the SWEET set, $\mathcal{S}_n$, due to both computational tractability, appropriateness for our Machine Learning (ML) technique, and its relevance for the targeted class of quantum states.

\subsection{Universal gates}\label{sec:universal_gates}
Any quantum circuit applied to qubit systems can always be built using a fixed finite set of gates or unitaries. This set of gates, say $\mathcal{U}$, is called universal gates.  $\mathcal{U}$ is not unique and can be modified according to available resources. 
In the context of quantum advantage, qubit gates can be classified into two groups, viz. Clifford and non-Clifford gates. Gates that map the Pauli group onto itself, i.e., the gates which normalize the Pauli group, are called Clifford gates. In addition to the Pauli gates themselves, Hadamard ($H$), {$\text{CZ}(c,t)$, and $\text{CNOT}(c,t)$} are also some examples of Clifford gates. Here 
\begin{equation}\label{eq_hadamard}
    H=\frac{1}{\sqrt{2}}\left(\begin{matrix}
        1&1\\
        1&-1
    \end{matrix}\right)
\end{equation}
and $\text{CZ}(c,t)$ ($\text{CNOT}(c,t)$) denotes the application of $Z$ ($X$) gate on the target qubit, $t$, when the control qubit, $c$, is in state $\ket{1}$, otherwise acting as the identity on $t$.

We know from the Gottesman–Knill theorem (\cite{gottesman1999heisenberg}) that circuits involving only Clifford gates can be simulated on classical computers. Therefore, to introduce computational quantumness in a circuit, one must add non-Clifford gates (gates that do not normalize the Pauli group) in the circuit, which suggests the inclusion of at least one non-Clifford gate in the set $\mathcal{U}$. One of the most popular single-qubit non-Clifford gates is 
\begin{equation}\label{eq_t_gate}
    T=\left(\begin{matrix}
        1&0\\
        0&e^{i\pi/4}
    \end{matrix}\right).
\end{equation}
To create entanglement between the qubits, we need at least one gate in the universal set that can produce entanglement, for example, the CNOT gate. Any set of gates that involves a class of gates that is dense in $\mathrm{SU}(2)$, one entangling gate, and one non-Clifford gate can be used as a universal set of gates. For building circuits to produce any $\ket{\psi_n}\in\mathcal{S}_n$, we will consider $\mathcal{U}=\{H,\text{CNOT},T, T^\dagger\}$ as the set of universal gates(\cite{Nielsen_Chuang_2010,Ga_MA_2002}). 

We introduce also an extra gate $T^\dagger$ for the following reason. Implementing a single $T^\dagger$ gate using only $T$ gates requires decomposing $T^\dagger$ into a sequence of seven $T$ gates, the circuit becomes unnecessarily long. Since preparing $T^\dagger$ directly is not intrinsically harder than preparing $T$, we decided to add directly a redundant gate rather than inflating the circuit depth by emulating $T^\dagger$ with multiple $T$ gates.

Therefore, in the rest of the paper, the notation $\mathcal{U}$ will be used to denote only this particular universal set of gates.

\subsection{Circuit complexity}
The circuit complexity of a quantum algorithm analyzes the difficulty in building and using that algorithm. Circuit complexity plays a crucial role in a wide range of fields (\cite{Brand_o_2021,watrous}). It serves as a useful tool for understanding the dynamics of quantum systems, especially in the context of chaos (\cite{Craps_2024}), decoherence (\cite{Haque_2022}), and phase transitions (\cite{Caputa_2022}).

The most traditional way of quantifying circuit complexity is by counting the total number of single- or two-qubit gates used in the circuit. However, circuit complexity can also be formulated in a continuous geometric framework (\cite{nielsen2005,Nielsen_2006}), where it is interpreted as the minimal path in the space of unitary operations needed to transform the identity into a given target unitary. Several approaches have been developed to describe this continuous notion of complexity (\cite{Jefferson_2017,Hackl_2018,Ali_2019}). One can also define circuit complexity of quantum algorithms concerning specific resources like entanglement or magic. In the following part, we discuss some of the circuit complexity measures that will be useful in analyzing the RL technique under consideration.
\begin{itemize}
    \item \textit{Gate count:} Gate count is the total number of gates used in the circuit. It is the most basic measure of circuit complexity. However, this measure provides equal weight to every gate, but in reality, particular gates can be more expensive than the others. That is why other measures, which are discussed below, are also useful. 
    \item \textit{$T$ count:} Non-Clifford gates are more expensive for fault-tolerant computation. Hence, the number of non-Clifford gates used in a circuit measures its complexity. In this work, we will use only one specific non-Clifford gate, i.e., the $T$ gate, and quantify the circuit complexity as the total number of $T$ gates used in the circuit. We name this particular type of complexity as $T$ count.
    \item \textit{Entangling-gate count:} Since entanglement is an essential quantum resource, the total number of entangling gates involved in the circuit also serves as a measure of complexity, which we refer to as entangling-gate count.
    \item \textit{Circuit depth:}  
    A quantum circuit may consist of gates that can be applied in parallel, specifically in situations where they operate on different parts of the entire system. The simultaneous action of such gates will reduce the execution time of the circuit, thus reducing the probability of having errors. We can divide the entire circuit into different time steps, where each time step can involve multiple gates that can be applied simultaneously. Circuit depth is defined as the total number of time steps in a circuit.
\end{itemize}

\section{Action Sequence-Based Q-Learning Using $\varepsilon$-greedy Exploration and Hybrid Rewards}\label{sec:3}
After defining and discussing all the necessary tools, in this section, we describe the reinforcement learning technique that we use to generate efficient quantum circuits.

\begin{figure}
    \centering
    \includegraphics[width=1.05\linewidth]{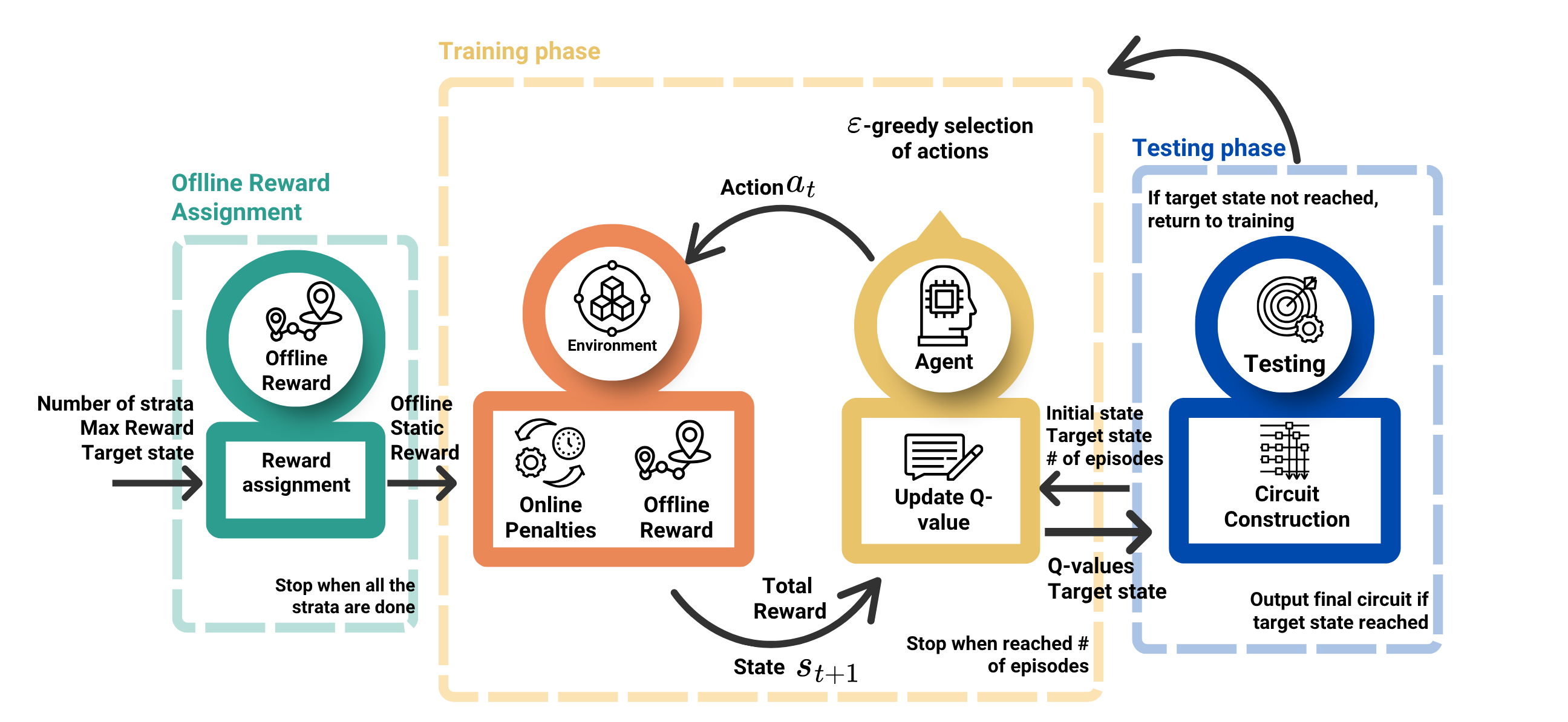}
    \caption{A pictorial representation of the overall pipeline that is illustrated in Section \ref{sec:3}. On the left, the green block is the offline reward assignment: given the target state, the number of strata, and the maximum value of the reward, we build an offline static reward map; this step runs once. At the center, the yellow block is the training phase, where an $\varepsilon$-greedy agent interacts with the environment; at each step it selects an action $a_t$, observes the next state $s_{t+1}$, and receives the total reward, which is fed back from the environment as a combination of the offline static reward and online penalties. Until the number of episodes is reached, the training continues. On the right, the blue block is the testing phase, with the learned $Q$ frozen, we construct with a completely greedy strategy the final circuit. If we reach the maximum allowed circuit length without having reached the target state, we return to the training phase. Otherwise, if we reach the target, we output the final optimized circuit. Arrows indicate the data flow between stages and the inputs and outputs of the three sections.}

    \label{fig:diagram}
\end{figure}

In Fig.~\ref{fig:diagram} there is a pictorial representation of the three main phases of the procedure: the construction of the offline reward, the training phase and the testing phase. These three parts will be explained in detail in the present section, together with the pseudocodes of their algorithms in Algorithm \ref{alg_reward}, Algorithm \ref{alg_training}, Algorithm \ref{alg_testing}.

\subsection{Q-learning basics}\label{subsec:q_learning_basics}
RL is used in machine learning to train an agent within a certain environment by giving it feedback based on its actions in the environment (\cite{sutton2018reinforcement}).
The RL technique that we use is called Q-learning, which can be used to teach the agent to find the best sequence of actions to perform in order to reach a specific state in the environment, without having any prior knowledge of the environment features. 

More specifically, in Q-learning, the agent has no prior knowledge of the goodness of an action $a$ performed in a specific state $s$, which implies that Q-learning is an off-policy RL algorithm (\cite{watkins1992q}). A policy is usually denoted as $\pi(s,a)$ and represents the value assigned to the pair action-state, based on the goodness of this choice, with respect to the target or goal. In Q-learning, the policy is usually represented by the Q-values $Q(s,a)$. This makes it different from other policy-based RL, which aims to optimize a preexisting policy by interacting with the environment. Thus, the objective is to define the best policy that will give the optimal path leading to a chosen final state from an initial state.

We can optimize different features of this path by choosing reward and penalty criteria and evaluating the values to feed back to the agent when it selects its actions. The reward function can be represented by a matrix, i.e., the R-matrix, which associates actions and states to a positive, negative, or zero value. This reward assignment can be done in an on-line or off-line fashion, by building a reward function $R(s,a)$, before the Q-learning starts or during the training process. 

The Q-learning procedure is divided into two parts. The first, the training phase, consists of exploring the environment, i.e., choosing a random action, $a_t$, and a random state, $s_t$, and building the quality matrix (Q-matrix) using the following Q-update rule:
\begin{gather}\notag
Q(s_t, a_t) \leftarrow Q(s_t, a_t) + \alpha \left[ R(s_t,a_t)  + \gamma \cdot \max_{a'} Q(s_{t+1}, a')
 -Q(s_t, a_t) \right], \label{eq_q_update}
\end{gather}
where $s_{t+1}$ denotes the state that the agent reaches by performing $a_t$ at $s_t$, $\alpha$ is the learning rate and $\gamma$ is the discount factor.
The Q-matrix finally specifies the goodness of the state-action pairs, thus the policy to follow to reach the desired state with the optimized path. 

The second part, the testing phase, consists of following the path indicated by the Q-values of the Q-matrix and checking if the learning process brings the agent to the desired target and if the obtained path respects the optimization criteria: a starting point is chosen, and the agent builds the chain of actions to perform in order to reach the final position, selecting in the Q-matrix the highest valued actions for each state it crosses. If it manages to reach the final position meeting the desired criteria for the path, then the learning process is considered successful, and the Q-matrix now represents a successful policy for the agent.

\subsection{Q-learning for circuit optimization: states and actions}\label{subsec:q_learning_for_circuits}
Let us now directly move to the problem at hand and discuss the Q-learning method in our specific circuit optimization context.

The states $s \in \mathcal{S}$ and actions $a \in \mathcal{A}$ that our agent evaluates correspond, respectively, to quantum states and quantum gates. We want to focus on the $n$-qubit SWEET set as described in Sec.~\ref{subsec:SWEET}.
To represent the discretized phase $\alpha_j$ of each term $j$ in the quantum superposition state $\ket{\psi_n}$, we introduce an auxiliary register containing $p$ additional qubits, supplementing the existing $n$-qubit system. Within this architecture, the $n$-qubit register encodes the computational basis state $\ket{x_j}$ that appears in the superposition $\ket{\psi_n}$. Since the phase $\alpha_j$ belongs to the set $\mathcal{P}_M$, it can be expressed in the form $\alpha_j = e^{i(2\pi m_j / M)}$, where $m_j$ is a natural number satisfying $0 \leq m_j < M = 2^p$.
The register of $p$ qubits registers the binary representation of the number $m$. For example, if $M=8=2^3$ then the state $\ket{00}-\ket{11}$ will be represented by $\ket{00000},\ket{10111}\}$ where the $3$ leftmost qubits for each term represent the phase $000 \to 1$ and $101 \to e^{i\pi}=-1$, while the $2$ rightmost qubits represent the real qubits. We would like to emphasize that there are no extra qubits involved in the circuit; we have introduced it only for the simulation of the discretized phases of the quantum states. This method is inspired by the sign-magnitude representation of binary numbers and allows us to implement gates that can alter the phases of the states and correctly store the result of the gate application, such as $T$ or $H$. Finally, we assign each of these SWEET states an index $s$. Since we have $p$ and $n$ qubits to register $\alpha_j$ and $\ket{x_j}$, respectively, the total number of forms that a particular term $\alpha_j\ket{x_j}$, of $\ket{\psi_n}$ can take is $2^{n+p}$. Since we consider that $j\neq j'$ implies $\alpha_j\ket{x_j}\neq \alpha_{j'}\ket{x_{j'}}$, the number of SWEET states having $N$ number of terms is $\binom{2^{n+p}}{N}$. Hence, the cardinality of the set, $\mathcal{S}_n$, of all SWEET states representing an $n$-qubit system is  
\begin{equation}\label{eq_N_S}
N_S = \sum_{N=1}^{2^{n+p}} \binom{2^{n+p}}{N}=2^{2^{n+p}}-1.
\end{equation}
Hence, the number of possible values that the index, $s$, which indicates each state, can take is $N_S$.
For example, when $n=7$ and $p=1$ (as we will consider to create graph states) $N_S=O(10^{77})$.

Meanwhile, the set of available actions $\mathcal{A}$ (see Sec.~\ref{subsec:q_training}) is defined as all valid applications of quantum gates from a fixed set over the $n$ qubits. Let us denote the number of distinct gates that jointly act on exactly $k$ qubits among $n$ qubits by $g^n_k$. For example, for $\mathcal{U} = \{H, T, T^\dagger, \text{CNOT}\}$, $g^n_1 = 3$ and $g^n_2 = 1$. 

The total number of valid actions associated with $k$-qubit gates is given by $A_k = g^n_k \cdot \text{Perm}(n, k)$, where $\text{Perm}(n, k) = \frac{n!}{(n - k)!}$ is the number of ordered $k$-tuples of distinct qubits that can be used as gate arguments (since the order matters for most gates, such as controlled gates). Thus, the total number of actions $N_A$ is:
\begin{equation}\label{eq_N_A}
N_A=|\mathcal{A}| = \sum_{k} g^n_k \cdot \text{Perm}(n, k).
\end{equation}
Hence, from the cardinality of the sets $\mathcal{A}$ and $\mathcal{S}_n$ we realize that the total dimension of the reward and quality matrices is $N_S\times N_A$. This expression grows exponentially with the number of qubits and the richness of the gate set. Combined with exponential growth in the state space, this further motivates the need for efficient exploration strategies and sparse Q-value storage.

Therefore, to store rewards and Q-values efficiently, using the least amount of memory, we use Structured Query Language (SQL) databases (\cite{Fundamentals_DB}). These are relational database systems designed to store and query structured data using tables. Each \emph{table} is defined by a set of \emph{columns} (fields) and stores data in \emph{rows}, where each row represents a distinct record or entry.

In our implementation, SQL is used to efficiently represent large, sparse matrices, such as $R$ and $Q$. Each nonzero element of these matrices, say $R(s,a)$ or $Q(s,a)$, is stored as a table of triplets as
\begin{equation}\label{eq_SQL}
\{\texttt{state\_index},\ \texttt{action\_index},\ \texttt{value}\}.
\end{equation}
This structure allows us to store only non-zero entries, significantly reducing memory usage. Notice that SQL databases are specifically efficient for queries such as reading or updating specific entries, which are the required actions for our algorithm.

Having in mind the initial state, $\ket{\psi_0}$, the target state $\ket{\psi_{\text{target}}}$, and the list of actions $\mathcal{A}$, we are ready to proceed.
\subsection{Static and dynamic reward}\label{subsec:r_static_dynamic}
We employ a hybrid reward strategy in which a sparse and static precomputed reward encourages the discovery of the shortest circuit reaching the target state, while a dynamic action-dependent reward function helps optimize secondary circuit properties such as preparation depth (\cite{rummery1994line}). This choice of using offline and online strategies helps us focusing on the environment exploration and avoid superfluous evaluations during online training of the algorithm, and aligns with the multi-objective reinforcement learning framework (\cite{roijers2013survey}) and the so-called "MaxPain" techniques (\cite{MaxPain}). Each element $R(s,a)$, of the reward matrix, is split as follows:
\begin{equation}\label{eq_reward_total}
R(s,a) = R_{\text{sta}}(s, a) + R_{\text{dyn}}(s, a).
\end{equation}
In Eq.~\eqref{eq_reward_total}, the static reward is precomputed and assigned as follows: 
\begin{equation}\label{eq_reward_static}
R_{\text{sta}}(s, a) =
\begin{cases}
R_{\text{max}} & \text{if } s \xrightarrow{a} s_{\text{target}} \\
\frac{R_{\text{max}}}{2}  & \text{if } \exists \ a, a_1 : s  \xrightarrow{a} s_1 \xrightarrow{a_1} s_{\text{target}} \\
\vdots & \vdots \\
\frac{R_{\text{max}}}{2^{k}} & \text{if } \exists \ a,a_1\dots a_{k-1}: s\xrightarrow{a,a_1\dots a_{k-1}} s_{\text{target}}\\
0    & \text{otherwise},
\end{cases}
\end{equation}
where $R_{\text{max}}>0$ is the maximum reward assigned and $s_{\text{target}}$ is the index representing $\ket{\psi_{\text{target}}}$. The other static rewards and penalty values are determined based on this reference value (\cite{wrro75111,sutton2018reinforcement}). To ensure that the algorithm functions effectively, penalties should not disrupt the main reward-based optimization. Therefore, individual penalties are set at approximately $1\%$ or $0.1\%$ of $R_{\text{max}}$. The reward layers help the agent reach the target state by creating a ``breadcrumb trail" for it.

To efficiently compute this static layered reward, we utilize the reversibility property of quantum gates(\cite{Nielsen_Chuang_2010}). Say $U$ is a quantum gate operator, then we have: $UU^{\dag}=U^{\dag}U=I$. We apply the Hermitian conjugate of a gate corresponding to column $j$, to the target state: $U_j^{\dag}\ket{\psi_{\text{target}}}=\ket{\psi_i}$ and fix the corresponding element of the static part of the R-matrix as $R_{\text{sta}}(s_i,a_j)=R_\text{max}$, where $s_i$ and $a_j$ represent the state $\ket{\psi_i}$ and the action of $U_j$ on $\ket{\psi_i}$, respectively. To construct the second layer, we apply $U_k^{\dag}$ to $\ket{\psi_i}$, resulting in the states for the second stratum $\ket{\psi_l}$ and set $R_{\text{sta}}[s_l, a_k] = R_{\text{max}} / 2$. This process is repeated for the remaining strata.
By repeating this procedure for all gates, we can build the reward matrix successfully. Here we write the steps of the reward assignment:
\begin{enumerate}
    \item Define the target state $\ket{\psi}_{\text{target}}$, the maximum reward value as $R_{\text{max}} = 10000$ and the number of strata $k_{\text{max}}$.
    \item Take $U_j$ from the available set of gates and compute $U^\dagger_j$.
    \item Apply $U^\dagger_j$ to the current target state $\ket{\psi_{\text{target}}}$, obtaining a new state $\ket{\psi_i}$.
    \item Let the current recursion strata be $k$. If $R_{\text{sta}}(s_i,a_j) < R_{\text{max}} / 2^k$, set $R_{\text{sta}}(s_i,a_j) = R_{\text{max}} / 2^k$.
    \item While $k < k_{\text{max}}$, apply steps 2–5 recursively with the updated state $\ket{\psi_{\text{target}}}\leftarrow\ket{\psi_i}$, $k \leftarrow k + 1$, and all possible gates $U_j$.
    \item Repeat for all possible gates $U_j$.
\end{enumerate}
After labeling each state, $\ket{\psi_i}\in \mathcal{S}_n$ with an index $s_i$, and each gate, $U_j$, acting on fixed qubits with an action $a_j$, we can assign rewards using the procedure outlined in Algorithm~\ref{alg_reward}.
\begin{algorithm}[t]
\caption{Recursive assignment of static reward values}
\begin{algorithmic}[1]
\State Set $s_{\text{target}}$
\State Set max reward $R_{\text{max}} \gets 10000$
\State Set maximum number of strata $k_{\max}$
\Procedure{AssignStaticReward}{$s_{\text{target}}, k$}
    \For{each action $U_j$ in the available gate set}
        \State Compute inverse $U_j^\dagger$
        \State Apply $a_j$ corresponding to $U_j^\dagger$ to $s_{\text{target}}$, and get $s_i$
        \If{$R_{\text{sta}}(s_i, a_j) < R_{\text{max}} / 2^k$}
            \State Set $R_{\text{sta}}(s_i, a_j) \gets R_{\text{max}} / 2^k$
        \EndIf
        \If{$k < k_{\max}$}
            \State $k \leftarrow k+1$
            \State $s_{\text{target}}\leftarrow s_i$
            \State \Call{AssignStaticReward}{$s_{\text{target}}, k$}
        \EndIf
    \EndFor
\EndProcedure
\end{algorithmic}\label{alg_reward}
\end{algorithm}
The dynamic component in Eq.~\eqref{eq_reward_total} is computed during Q-learning training, based on the sequence of actions taken in each episode and their varying impacts on the environment. It involves penalties assigned to the agent for following ineffective paths. Specifically, penalties are applied for:
\begin{enumerate}[i.]
    \item revisiting states within the same episode,
    \item performing actions that do not change the state,
\end{enumerate}
In addition, we can incorporate other penalties based on specific requirements. For instance, penalties can be added to discourage the use of costly gates or to reduce the circuit depth. These specific examples are discussed in the next section.

During each training episode, to impose penalties for i. revisiting states and ii. taking ineffective actions, we record the visited states in a list called $V$. We check $V$ to identify revisited or unchanged states (see Eq.\eqref{eq_reward_penalty}). These penalties are directly incorporated into the reward signal and included in the standard Q-learning update.

To set magnitudes, we fix the higher layers of reward at $R_{\max}\!\approx\!10^4$ and assign successive strata $R_{\max}/2,\,R_{\max}/4,\,R_{\max}/8,\ldots$. With discount factor $\gamma=0.5$, the discounted value of eventually collecting $R_{\max}$ after $d$ steps is
\begin{equation}
R_{\max}\gamma^d \;=\; 10^4 \cdot 0.5^d.
\end{equation}
If our per-step penalty is of size $10$ then, we equate the discounted reward to this penalty value $10^4\cdot 0.5^d\!\approx\!10$ getting the number of steps in which penalties and discounted rewards are equal $\Rightarrow d\!\approx\!10$; equating it to $1$ yields $d\!\approx\!13$. Thus, beyond roughly $10$--$13$ steps from the target, the future gain becomes numerically comparable to a single penalty. Since each step corresponds to one gate of the circuit, this matches our regime of interest, with few qubits and circuits under $10$--$15$ gates. Near the target, layered rewards dominate so the agent is not discouraged from exploring by the penalties; far from any good path, penalties are already strong enough to suppress unproductive moves. We therefore tuned the ratio between $R_{\max}$ and penalty magnitudes so that this crossover falls within the expected circuit lengths for our benchmarks (e.g., graph-state preparations), thus the penalties will be $\approx R_{\max}\cdot10^{-4}$ or $R_{\max}\cdot10^{-3}$.

\begin{equation}\label{eq_reward_penalty}
R_{\text{dyn}}(s_t, a_t ) = 
\begin{cases}
R_{\text{dyn}}(s_t,a_t)-R_{\text{max}}\cdot10^{-4} & \text{if } s_{t}\xrightarrow{a_t} s_{t+1}\; \; \text{where } s_{t+1}\in V \\
& \qquad \qquad \qquad \; \text{and } R_{\text{sta}}(s_t, a_t) = 0 \\ 
R_{\text{dyn}}(s_t,a_t)-R_{\text{max}}\cdot10^{-3} & \text{if } s_{t}\xrightarrow{a_t} s_{t+1}=s_t.\\
\end{cases}
\end{equation}
The absolute value $R_{\max}\!\approx\!10^4$ is not physically special, what matters is the separation of scales and the ratio between layers and penalties, in line with established RL control logic where the primary objective receives the largest weight and secondary terms enter with much smaller weights (see, \cite{Bukov2018,Niu2019UniversalControl}). For larger systems, where valid circuits may require substantially more steps, this balance should be revisited, either by increasing $R_{\max}$ or by adjusting the reward/penalty ratio, so that penalties continue to guide exploration without ever overriding the drive to reach the target once a viable path is found.

After outlining the process of constructing the reward matrix, we now move on to the details of the agent training method.

\subsection{Training phase for action-sequence based Q-learning}\label{subsec:q_training}
Our approach utilizes the standard tabular Q-learning algorithm, which iteratively updates the value function $Q(s,a)$ through temporal difference (TD) learning and the traditional Q-update rule. This is achieved by randomly selecting gates to apply to randomly chosen states and exploring the entire state-action space (\cite{sutton2018reinforcement, RusselAI}).

The training process is organized into episodes. Each episode begins with the agent starting from a fixed initial state $s_{0}$, representing a specific quantum state $\ket{\psi_0}\in\mathcal{S}_n$. 
The agent then performs a sequence of $k$ actions, updating the state after each action. Notably, in this setup, the initial state is not randomly chosen, as in standard Q-learning.
At each step, the agent can select an action either randomly from the set of possible actions $\mathcal{A}$ (exploration) or by choosing the action with the highest current Q-value (exploitation). To balance exploration and exploitation, we use the $\varepsilon$-greedy strategy (\cite{sutton2018reinforcement}): with probability $\epsilon$, the agent explores by selecting a random action, and with probability $1-\epsilon$, it exploits by choosing the best-valued action.
This sequential process maintains the off-policy nature of Q-learning while promoting exploration and learning in high-dimensional environments.

\begin{algorithm}[t]
\caption{Training phase of the Q-learning algorithm}
\begin{algorithmic}[1]
\State Define learning parameters: exploration rate $\epsilon$, learning rate $\alpha$, discount factor $\gamma$
\State Initialize the Q-matrix
\State Set number of episodes $l_e$ and episode length $k$
\State Define the set of possible actions $\mathcal{A}$
\For{each episode $e = 1$ to $l_e$}
\Statex \Comment{Fixed initial SWEET state}
    \State Initialize $s_0$ corresponding to $\ket{\psi_0}$
    \For{each step $t = 0$ to $k - 1$}
        \State Select action $a_t \in \mathcal{A}$ using $\varepsilon$-greedy strategy:
        \State with probability $\epsilon$, choose $a_t$ randomly 
        \Statex \Comment{Exploration}
        \State with probability $1 - \epsilon$, choose $a_t = \max_{a} Q(s_t, a)$
        \Statex \Comment{Exploitation}
        \State Apply $a_t$ to $s_t$, obtaining new state $s_{t+1}$
        \State Compute $R(s_t,a_t) = R_{\text{sta}}(s_t, a_t) + R_{\text{dyn}}(s_t, a_t)$
        \State Update Q-value using temporal difference:
        \begin{equation*}
            Q(s_t, a_t) \leftarrow Q(s_t, a_t) + \alpha \left[R(s_t,a_t) +  \right. 
        \end{equation*}
        \begin{equation*}
            \left.+ \gamma \cdot \max_{a'} Q(s_{t+1}, a') - Q(s_t, a_t)\right]
        \end{equation*}
        \State Set $s_t \leftarrow s_{t+1}$
    \EndFor
\EndFor
\State Proceed to testing phase (see Algorithm~\ref{alg_testing})
\end{algorithmic}
\label{alg_training}
\end{algorithm}

We denote the sequence of actions in one episode by $A$. After completing $l_e$ episodes of fixed length $k$, the algorithm transitions to the testing phase. The values $k$ and $l_e$ depend on the specific application. Training continues with batches of episodes until the testing phase proves successful (see Sec.~\ref{subsec:q_testing}).
The primary objective of training is to iteratively build and update the Q-matrix, $Q$, which represents the expected long-term reward for taking action $a$ in state $s$. During each episode, after every applied action, the agent receives a reward signal composed of both static and dynamic components, as detailed previously in Eq.~\eqref{eq_q_update} and Sec.~\ref{subsec:r_static_dynamic}. The static reward, computed offline, directs the agent toward the target state by rewarding minimal length paths and is stored as a sparse matrix. The dynamic reward is calculated online during training, with the elements of the matrix $R_{\text{dyn}}(s,a)$ constructed according to Eq.~\eqref{eq_reward_penalty}.

These reward signals are used to update the Q-values using the temporal difference learning rule in Eq.~\eqref{eq_q_update}, where $R(s_t,a_t)$ represents the total reward obtained at step $t$, as described in Eq.~\eqref{eq_reward_total}. Across multiple episodes, the Q-matrix progressively encodes the most efficient action sequences for reaching the target quantum state, optimizing not only for gate count but also for any additional criteria if specified.

Our training structure aligns with Q-learning based on action sequences and planning-inspired exploration, similar to models like Dyna-Q (\cite{sutton1991dyna, sutton1990integrated, blundell2016model}). However, our algorithm remains strictly off-policy and tabular: it does not utilize neural network approximators as in Deep Q-learning (\cite{Mnih2015}), nor does it directly optimize a policy as in policy gradient or actor-critic methods (\cite{sutton2000policy, konda2000actor}). This setup enables more efficient exploration in high-dimensional spaces while still allowing Q-learning to converge to an optimal solution under suitable exploration strategies.

The training phase occurs before the testing phase (see Sec.~\ref{subsec:q_testing}), during which the effectiveness of the algorithm is evaluated. The training process is repeated in batches until satisfactory performance is achieved in the testing phase. This sequential approach, together with the use of hybrid rewards and a discretized representation of quantum states, facilitates effective learning in high-dimensional and sparse environments.

\subsection{Q-learning testing phase}\label{subsec:q_testing}
To evaluate the learned policy matrix $Q$, we generate a circuit based on $Q$ after completing a batch of episodes and then assess the training success. The testing phase begins by selecting the row $s_0$, representing the initial state $\ket{ \psi_0}$, and setting a maximum circuit length $l_c$. We identify the action $a_0$ for which $Q(s_0, a_0) \geq Q(s_0, a)$ for all $a$. The corresponding gate $U_0$, represented by $a_0$, is applied to $\ket{\psi_0}$, resulting in $\ket{\psi_1} = U_0 \ket{\psi_0}$. We then check if $\ket{\psi_1}$ matches the target state $\ket{\psi_{target}}$. If not, we find the row $s_1$ corresponding to $\ket{\psi_1}$ and repeat the process: selecting the action $a_t$ that maximizes $Q(s_t, a)$, applying the associated gate $U_t$, and updating the state to $\ket{\psi_{t+1}}$. This cycle continues until the state matches the target $\ket{\psi_{\text{target}}}$, or until the maximum circuit length $l_c$ is reached (i.e., $t\leq p$). If successful, the algorithm concludes; otherwise, the process returns to training.

Hence, the testing phase is as follows.
\begin{enumerate}
    \item Find the row $s_i$, which represents the state $\ket{\psi_i}$.
    \item Find the column $a_i$, for which $Q(s_i,a_i)\geq Q(s_i,a)$ for all $a$.
    \item Apply the gate, $U_i$, represented by the column, $a_i$, on $\ket{\psi_i}$. 
    \item Check if $U\ket{\psi_i}$ equals the target state. If it does, the testing phase terminates. If not and the total circuit length is less than $l_c$, reset $\ket{\psi_i}$ to $U_i\ket{\psi_i}$ and return to Step 1.
    \item If the circuit length reaches $l_c$, the process switches back to the training phase, initiating a new training batch. The Q-matrix used for testing is used to initialize the Q-matrix for this new training batch.
\end{enumerate}

After the algorithm concludes, the sequence of gates corresponding to the actions $\{a_0,\dots,a_{t-1}\}$ constitutes the final circuit. This circuit transforms the initial state $\ket{\psi_0}$ into the target state $\ket{\psi_{\text{target}}}$ using the minimum number of gates, which is less than $l_c$, the maximum circuit length specified at the start. 

This Q-learning approach not only constructs a circuit with an optimal length but also allows the incorporation of additional constraints through suitable penalties. For instance, we can discourage the use of T-gates, entangling gates, or any specific expensive gate by updating the dynamic component of the reward matrix with penalties. Additionally, appropriate penalties can be employed to minimize circuit depth. These optimization strategies will be discussed in more detail in Sec.~\ref{sec:4}.

\begin{algorithm}[t]
\caption{Testing phase of the Q-learning algorithm}
\begin{algorithmic}[1]
\State Set initial state $s_0 $ corresponding to $ \ket{\psi_0}$
\State Set target state $s_{\text{target}} $ corresponding to $ \ket{\psi_{\text{target}}}$
\State Set maximum allowed circuit length $l_c$
\State Initialize step counter $t \gets 0$
\While{$s_t \neq s_{\text{target}}$ \textbf{and} $t < k$}
    \State Select action $a_t = \max_{a} Q(s_t, a)$ 
    \State Apply gate corresponding to $a_t$ to $s_t$, and obtain $s_{t+1}$
    \State Set $s_t \leftarrow s_{t+1}$
\EndWhile
\If{$s_t = s_{\text{target}}$}
    \State Sequence $\{a_0, a_1, \dots, a_{t-1}\}$ defines the final circuit
\Else
    \State Testing failed: maximum circuit length $l_c$ reached:
    \Statex return to training phase in Algorithm~\ref{alg_training}
\EndIf
\end{algorithmic}
\label{alg_testing}
\end{algorithm}

\subsection{Exploration complexity and model parameters}
\label{subsec:exploration_complexity}

The exploration space generated by our formulation is naturally large because of the combinatorial complexity of the quantum states involved. Given that $n$ and $p$ denote the number of qubits and phase qubits used to encode the states, respectively, the full Hilbert space is defined over $n+p$ qubits. This results in a computational basis $\mathcal{B}_{n+p}$ of size $2^{n+p}$. The total number of states, $N_S$, that the agent explores, i.e. the cardinality of $\mathcal{S}_n$, is even greater.

Within this vast space, the reward is highly sparse, making exploring the environment particularly challenging. To address this, we do not explicitly store or explore the entire state-action space. Instead, we utilize sparse representations of the Q-matrix and the R-matrix in database-style structures of the form \{state index, gate index, value\}, where each entry represents a valid (nonzero) Q-value or reward, as described in Sec.~\ref{subsec:q_learning_for_circuits}. This approach allows us to efficiently track only the explored or relevant transitions, significantly reducing memory consumption and computational overhead.

In the training part we chose a high exploration rate ($\varepsilon=0.8$) in our $\varepsilon$-greedy strategy, to efficiently probe a large, sparsely rewarded state space. This increases the capability of the algorithm of covering a large part of admissible states and helps the agent to reach the layered rewards rather than falling into local loops. This choice is aligned with tabular Q-learning being off-policy, while negative penalties drive the exploration away from useless regions. In the testing part the agent acts in a fully greedy (pure-exploitation) mode.

We set the discount factor to $\gamma=0.5$ and the learning rate to $\alpha=0.8$. The discount factor controls the effective planning horizon in the Q-learning target (see Eq.\eqref{eq_q_update}. With layered positive rewards near the target and strictly negative per-step penalties, $\gamma=0.5$ focuses learning on the short-to-medium horizon that is relevant for few-qubit graph-state circuits, while attenuating remote returns. In particular, the discounted contribution of a terminal reward $R_{\max}$ after $d$ steps is $R_{\max}\,0.5^{\,d}$, which becomes comparable to a single penalty ($\sim\!1$–$10$) only after about $d\!\approx\!10$–$13$ steps; this aligns the update signal with the typical circuit depths considered. The high learning rate $\alpha=0.8$ helps not to lose the sparse rewards once encountered, since the learning rate is responsible to propagate the reward signal learned in each step.

Furthermore, employing sequence-based Q-learning with a fixed initial state and structured penalties promotes efficient exploration and helps the agent to reliably converge to the target state (\cite{watkins1992q}). The dynamic penalties--designed to discourage unnecessary actions--stop the agent from aimless wandering and speed up the learning process.

However, the method is still affected by the curse of dimensionality (\cite{dulac2019challenges}). Although representing the Q-table and the reward matrix as sparse matrices greatly reduces memory requirements and improves storage scalability (\cite{li2009sparse}), the sparsity of static rewards presents a different challenge. As the number of qubits grows, the likelihood of reaching a rewarding state-action pair, where $R(s,a)>0$, decreases quickly. To address this, we implement static reward strata at progressively greater distances from the target state.

To address this limitation and keep training computationally manageable, the examples presented in the following sections are limited to systems with up to $n=7$, $p=1$ qubits, and $n=5$, $p=3$.

\section{Implementation and results}\label{sec:4}

\begin{figure}[t]
   \centering
   \includegraphics[width=0.8\linewidth]{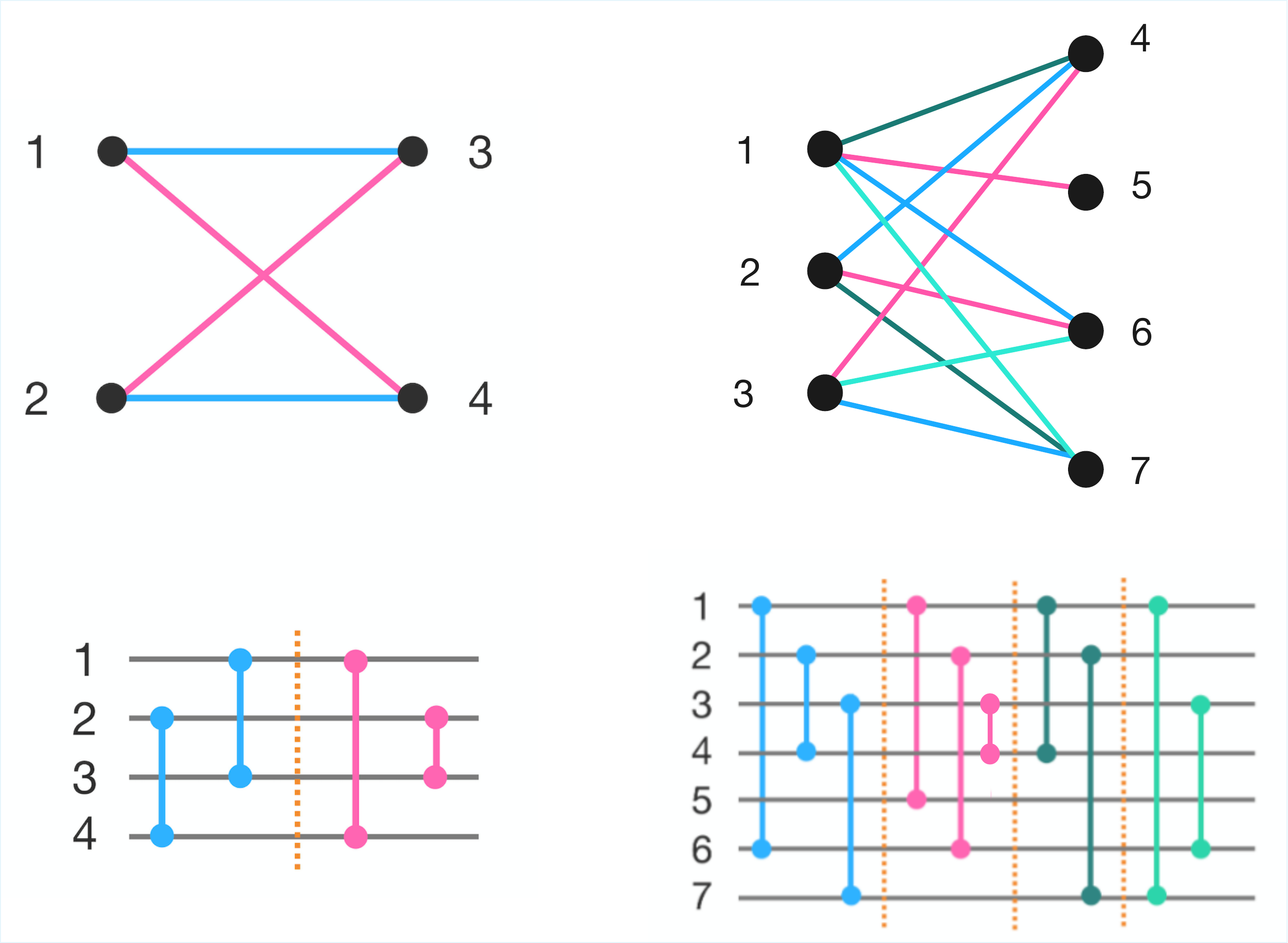}
   \caption{The figure illustrates the structure of two specific graphs alongside diagrams of the circuits used to generate them. The top panel displays a $4$-vertex graph (left) and a $7$-vertex graph (right). Edges of the same color signify that the corresponding CZ gates can be applied simultaneously at identical time steps to create the graph states. The bottom panel shows the optimal circuits with minimal depths required to generate these graph states. In the circuit diagrams, vertical solid lines represent the application of CZ gates, while colored circles indicate the qubits involved in each gate's application. Vertical dashed lines separate different time steps.}
   \label{fig:Graph_states_results}
\end{figure}

After reviewing the algorithm details, this section demonstrates how the proposed tabular Q-learning framework converges toward high-quality solutions and how this convergence materializes in concrete quantum circuits. We proceed to illustrate first a benchmarking of the algorithm on graph states, and second a more general result employing a universal set of gates.

\subsection{Benchmarking results on graph states}\label{sub:benchmarking_intro}

We notice that the chosen SWEET state set is well suited for representing graph states (\cite{RaussendorfGraphStates}). Graph states form a key class of highly entangled quantum states, widely utilized as resources in measurement-based quantum computation (\cite{hein2006graphstates}). Their structured entanglement also supports applications in quantum error correction, communication, and multipartite entanglement research (\cite{multiparty_entanglement_graph_states}). Before analyzing the performance of the learning algorithm, we first clarify the formal structure of graph states and discuss the theoretical background, which makes them ideal benchmarking candidates.

%\subsection{Graph states definition}\label{subsec:graph-structure}
A graph, $G=(\mathcal{V},\mathcal{E})$, is defined through a set $\mathcal{V}$ of vertices (or nodes) and a set of edges, $\mathcal{E}\equiv\{\{v_i,v_j\}|v_i,v_j\in\mathcal{V}\}$. For every graph having $n$ vertices, we can always define a corresponding $n$-qubit state as:
\begin{equation}\label{eq_graph_states}
    \ket{G_n}=\prod_{\{v_i,v_j\}\in\mathcal{E}}\text{CZ}( v_i,v_j)\ket{+}^{\otimes n},
\end{equation}
hence, called graph state. For an $n$-qubit computational basis state $\ket{x_1 x_2 \dots x_n}$ with $x_k \in \{0,1\}$, the action of the CZ gate on qubits $i$ and $j$ is defined as:
\begin{equation}\label{eq_cz}
\text{CZ}(i,j) \ket{x_1 x_2 \dots x_n} =
(-1)^{x_i x_j} \ket{x_1 x_2 \dots x_n}.
\end{equation}
That is, the state acquires a phase of $-1$ only if both $x_i = 1$ and $x_j = 1$.  Thus, $\text{CZ}(v_i,v_j)$ is the gate $\text{CZ}$ acting on the qubit corresponding to vertex $v_j$ and controlled by the qubit corresponding to vertex $v_i$. The initial state used for graph state preparation is $\ket{+}^{\otimes n}$, a uniform superposition of the computational basis, and we denote it as $\ket{\psi^{in}_{n}}$. Therefore, the number of CZ gates required to create a graph state from $\ket{\psi_n^{in}}$ is known and equal to the total number of edges of the graph. CZ gates entangle qubits according to the edges of $G$. Since CZ gates commute, the total number of CZ gates equals $|\mathcal{E}|$, and the execution order can be rearranged to minimize circuit depth. Graph states can be seen as entangled stabilizer states whose structure is fully determined by the connectivity of the underlying graph.

\subsection{Verification against theoretical optima}\label{subsec:benchmark_setup}
We now describe the specific target states used to benchmark the reinforcement learning algorithm, along with their known theoretical depth bounds. We will focus on the graph states corresponding to the graphs depicted in Fig.~\ref{fig:Graph_states_results}, $\ket{G_4}$ and $\ket{G_7}$.

Let us define as $\mu(\ket{\psi},\ket{\phi})$ the number of gates composing a circuit needed to generate state $\ket{\phi}$ starting from state $\ket{\psi}$. To create the graph states $\ket{G_4}$ and $\ket{G_7}$ that have 4 and 7 vertices, and 4 and 10 edges, respectively, we need $\mu(\ket{\psi_4^{in}},\ket{G_4})=4$ and $\mu(\ket{\psi_7^{in}},\ket{G_7})=10$ CZ gates. 

Let us also define, analogously to $\mu(\ket{\psi},\ket{\phi})$, the depth of the circuit generating $\ket{\phi}$ from $\ket{\psi}$ as $\delta(\ket{\psi},\ket{\phi})$. Using the Vizing's theorem (\cite{MISRA1992131}), it can be shown that the minimum depth, $\delta(\ket{\psi_n^{in}},\ket{G})$, of the circuit that generates a graph state $\ket{G}$ using CZ gates, starting from $\ket{\psi_n^{in}}$, is directly related to the maximum degree, $\Delta(G)$, of the graph, $G$. The degree of a vertex is the number of edges connected to the vertex. In particular, $\delta(\ket{\psi_n^i},\ket{G})$ is either $\Delta(G)$ or $\Delta(G)+1$. For bipartite graphs (like those depicted in Fig.~\ref{fig:Graph_states_results}) the bound tightens to $\Delta(G)$. 

Because for these bipartite graphs the edge–chromatic number equals the maximum degree, $\chi'(G)=\Delta(G)$ (by Vizing’s theorem), any CZ schedule with depth $\Delta(G)$ is provably optimal; thus matching this depth constitutes a verification against a known theoretical optimum. Prior work has been conducted on optimized circuits for graph-state preparation using combinatorial and constructive methods (e.g., \cite{Cabello2011,Liu2023,kumabe2024}), but to the best of our knowledge reinforcement-learning–based approaches targeting optimal graph-state circuits have not been reported yet.

Hence:
\begin{align*}
    \delta(\ket{\psi_4^{in}},\ket{G_4}) &= 2, \\
    \delta(\ket{\psi_7^{in}},\ket{G_7}) &= 4.
\end{align*}
These known values will allow us to evaluate whether the learning agent can match the theoretical optimal circuits in terms of both depth and gate count, in particular for states $\ket{G_4}$ and $\ket{G_7}$. The initial states for the circuit generation are $\ket{\psi_4^{in}}$ and $\ket{\psi_7^{in}}$, respectively. The SWEET sets that we focus on are $\mathcal{S}_4$ and $\mathcal{S}_7$ (defined in Subsection ~\ref{subsec:SWEET}) and the actions considered, $\mathcal{A}$, correspond to CZ gates. We specify that the phase set we are considering is $\mathcal{P}_2=\{1,-1\}$, i.e., we take $M=2$ in the SWEET states definition in Eq.~\eqref{eq_SWEET}. $N_S$ and $N_A$ can be obtained using Eqs.~\eqref{eq_N_S} and \eqref{eq_N_A} by putting $n=$4 and 7 respectively for $\ket{G_4}$ and $\ket{G_7}$, and considering $g^n_1=0$ and $g^n_2=1$, as defined in Sec.~\ref{subsec:exploration_complexity}.

\subsection{Reward strata and penalty design}\label{subsec:reward_and_pen_for_graph_states}
The RL setup described in Sec.~\ref{sec:3} is tailored here to favor low-depth circuit discovery for graph states. Specifically, to minimize circuit depth, we define a \emph{congestion penalty} that penalizes actions that reuse qubits already involved in recent gates.

We evaluate how much the qubits involved in an action $a_t$ have been involved in the past actions of the same episode sequence, defining the ``congestion level" $\text{C}(a_t)$ as follows. First, we define a counter vector $\mathbf{c} = (c_1, \dots, c_n)$, initialized at the beginning of each episode as $c_i = 0 \quad \text{for all } i \in \{1, \dots, n\}$. The size of the vector is carefully chosen to be equal to the number of qubits $n$ involved in the target state. We iterate over all the actions prior to action $a_t$ in the episode, i.e. $A_{<t} = \{a_0, \dots, a_{t-1}\}$. For each qubit $q_i$ used by $a_j \in A_{<t}$ we increment the corresponding counters $ c_i \leftarrow c_i + 1$. Considering that the current action $a_t$ acts on a subset $\mathcal{Q}_{a_t}$ of qubits, the congestion level is:
\begin{equation}\label{eq_congestion_level}
    \text{C}(a_t) = \underset{q_i \in \mathcal{Q}_{a_t}}{\max} c_i.
\end{equation}
We then apply the penalty:
\begin{equation}\label{eq_depth_penalty}
R_{\text{dyn}}(s_t, a_t ) = R_{\text{dyn}}(s_t, a_t ) -R_{\text{max}}\cdot10^{-4} \quad \text{if } \text{C}(a_t) > \text{CT}_t,
\end{equation}
where $\text{CT}_t = t/2$ is a congestion threshold depending on the time step $t$. This mechanism discourages serialized actions and encourages parallel gate application.

\subsection{Training phase and visualization of the learning process}\label{subsec:rl_visual}
\begin{figure}[t]
   \centering
   \includegraphics[width=1\linewidth]{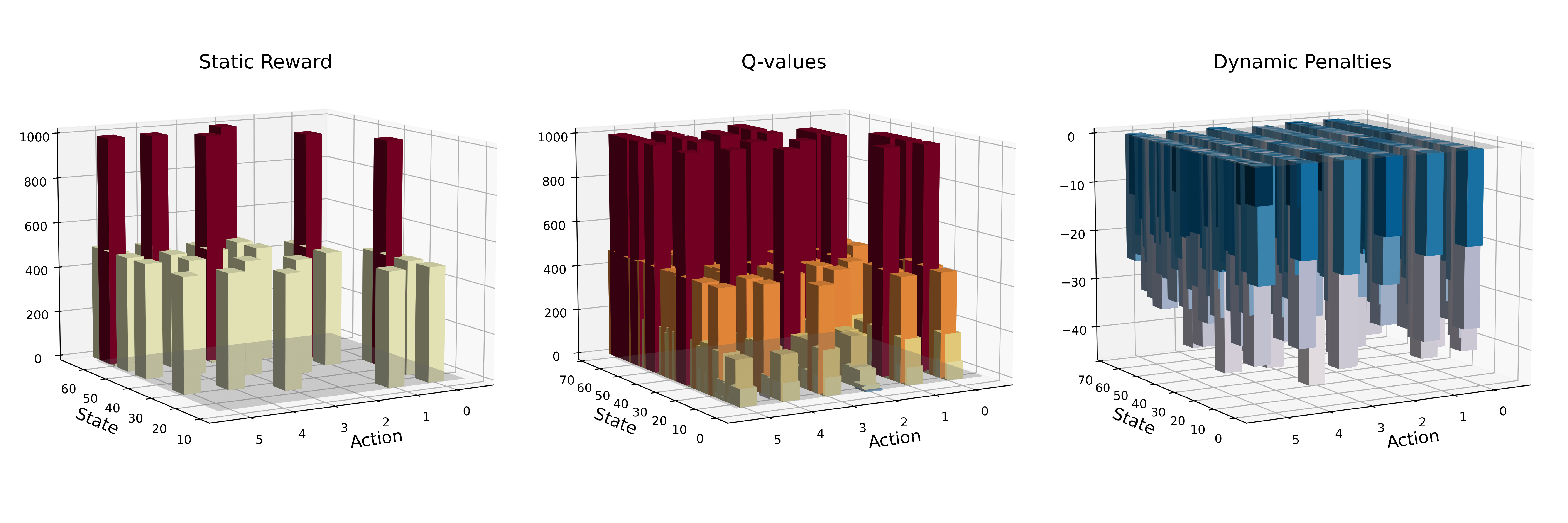}
   \caption{The learned Q-matrix and its constituent reward components. The figure presents 3D bar charts of the matrices used in the Q-learning algorithm for a $4$-qubit graph state. From left to right, the first chart depicts the static reward matrix $R_{\text{sta}}(s,a)$ with two reward strata. The second shows the learned Q-values $Q(s,a)$ after training. The third illustrates the dynamic penalty matrix $R_{\text{dyn}}(s,a)$, computed online. Color gradients indicate the magnitude and sign of values: warm colors for positive and cool colors for negative. For the purpose of clear visualization, only the rows and columns containing nonzero entries of the matrices are shown. Collectively, these plots illustrate how static guidance and dynamic penalties interact to construct and learn a policy. For more details, see Sec.~\ref{subsec:r_static_dynamic} and Subsection~\ref{subsec:rl_visual}.
   }
   \label{fig:Q_learning_plots_4_qubits}
\end{figure}

In this subsection, we analyze the learning dynamic summarized in Fig.~\ref{fig:Q_learning_plots_4_qubits} which brings to the optimal circuits of Fig.~\ref{fig:Graph_states_results}.

In particular, Fig.~\ref{fig:Q_learning_plots_4_qubits} condenses the entire training history for creating $\ket{G_4}$ into three sparse matrices, the static reward matrix, $R_{\text{sta}}(s,a)$ (left panel), the learned Q-values $Q(s,a)$ (center panel), and dynamic penalty map $R_{\text{dyn}}(s,a)$ (right panel). Notice that, to facilitate interpretation, the plot includes only state-action pairs with non-negative values.

The left panel in Fig.~\ref{fig:Q_learning_plots_4_qubits} encodes the layered breadcrumb trail introduced in Sec.~\ref{subsec:r_static_dynamic}. Only state--action pairs included in the two reward strata described in Eq.~\eqref{eq_reward_static} have nonzero entries, making the figure visibly sparse. The tallest bars ($R_{\max}=10^{4}$) correspond to single--gate transitions that reach the target state directly, while the shorter bars at $R_{\max}/2$ correspond to states that are two action-steps away. For both states $\ket{G_4}$ and $\ket{G_7}$ we set $R_{max}=10^{4}$ and $k_{\text{max}}=2$, to have two reward strata.

The central panel in Fig.~\ref{fig:Q_learning_plots_4_qubits} shows nonzero $Q$ bars highlighting the spread reward with smoothly varying amplitudes. This amplification results from temporal difference updates (see Eq.~\eqref{eq_q_update}). We recall that we considered the following values of the training parameters for both $\ket{G_4}$ and $\ket{G_7}$: $\epsilon=0.8$, $\alpha=0.8$, $\gamma=0.5$ for the $\varepsilon$-greedy strategy and the Q-learning update, respectively, see Section\ref{subsec:exploration_complexity} for more details on the choice of parameters. 

The algorithm provided the circuits in Fig.~\ref{fig:Graph_states_results} within a training of $10^{4}$ episodes for $\ket{G_4}$ and $7\times 10^{4}$ for $\ket{G_7}$, where each episode involves action sequences of length 50 and starts from the initial state $\ket{\psi_4^{in}}$ and $\ket{\psi_7^{in}}$, respectively.

The right panel shows $R_{\text{dyn}}$, containing exclusively negative values due to penalties applied for revisiting states, actions leaving the state unchanged, or exceeding congestion thresholds. The distribution of cool-colored bars confirms that exploration is steered away from inefficient behaviors.

Taken together, the panels confirm convergence: the Q-table inherits the sparse structure of $R_{\text{sta}}$ and avoids high-penalty regions in $R_{\text{dyn}}$. The training hyperparameters were empirically tuned to balance exploration and convergence stability. 

\subsection{Consideration on average Bellman error}\label{subsec:bellman_error}
In Q-learning, the Bellman error quantifies the discrepancy between the current estimate of the Q-value for a state-action pair and the updated value predicted by the Bellman equation shown in Sec.~\ref{subsec:q_learning_basics}. Minimizing this error is fundamental for the algorithm to converge to the optimal Q-table and produce accurate value estimates, and is the result expected after a successful training phase (\cite{sutton2018reinforcement,Csaba, Mnih2015}). Specifically, the Bellman error at each update step is given by the temporal difference term:
\begin{equation}\label{eq_bellman_error}
\delta_t = R(s_t,a_t) + \gamma \cdot \max_{a'} Q(s_{t+1}, a') - Q(s_t, a_t).
\end{equation}
By iteratively minimizing $\delta_t$, the algorithm refines its Q-values towards the true expected returns, thus learning an optimal policy.

At the end of the training phase of our algorithm, we measure an average Bellman error well below 1\% of the maximum reward scale $R_{\max} = 10^{4}$, typically below 100 in absolute value. It is important to stress that this 1\% threshold is a practical heuristic rather than a universal theoretical criterion (\cite{Csaba, Mnih2015}). Such a low average Bellman error confirms that the learned Q-values closely approximate the expected future rewards, signaling effective convergence of the Q-learning process. This precision level aligns with common practical heuristics in reinforcement learning. Consequently, the small Bellman error reinforces confidence in the robustness and near-optimality of the quantum circuits synthesized by the learned policy.

\subsection{Testing phase and circuit reconstruction}
\label{subsec:testing_phase}
{To validate the learned policy, we run the testing phase (Algorithm~\ref{alg_testing}). Starting with the same initial state of the episodes, the action with the maximum Q-value is selected until the target state is reached or the circuit length cap $l_c = 50$ is reached.}

Fig.~\ref{fig:Graph_states_results} shows the final circuits obtained for the two benchmark graphs:
\begin{itemize}
    \item Four-qubit square graph ($\ket{G_4}$): A depth-2 circuit with 4 CZ gates.
    \item Seven-qubit bipartite graph ($\ket{G_7}$): A depth-4 circuit with 10 CZ gates. 
\end{itemize}
These results match the optimal theoretical bounds derived from Vizing's theorem (\cite{MISRA1992131}): $\delta(\ket{\psi_4^{in}},\ket{G_4})=\Delta(G_4)=2$ and $\delta(\ket{\psi_7^{in}},\ket{G_7})=\Delta(G_7)=4$. Thus, the Q-table, shaped by static rewards and dynamic penalties, encodes the parallelizable structure of these circuits. The learned policy thus yields not only gate-optimal but also discovers the most parallelized structure permitted by the target state, thus attaining depth-optimal results.

\subsection{Scalability remarks for graph states}\label{subsec:scalability_remarks}
\begin{figure}[t]
    \centering
    \includegraphics[width=1\linewidth]{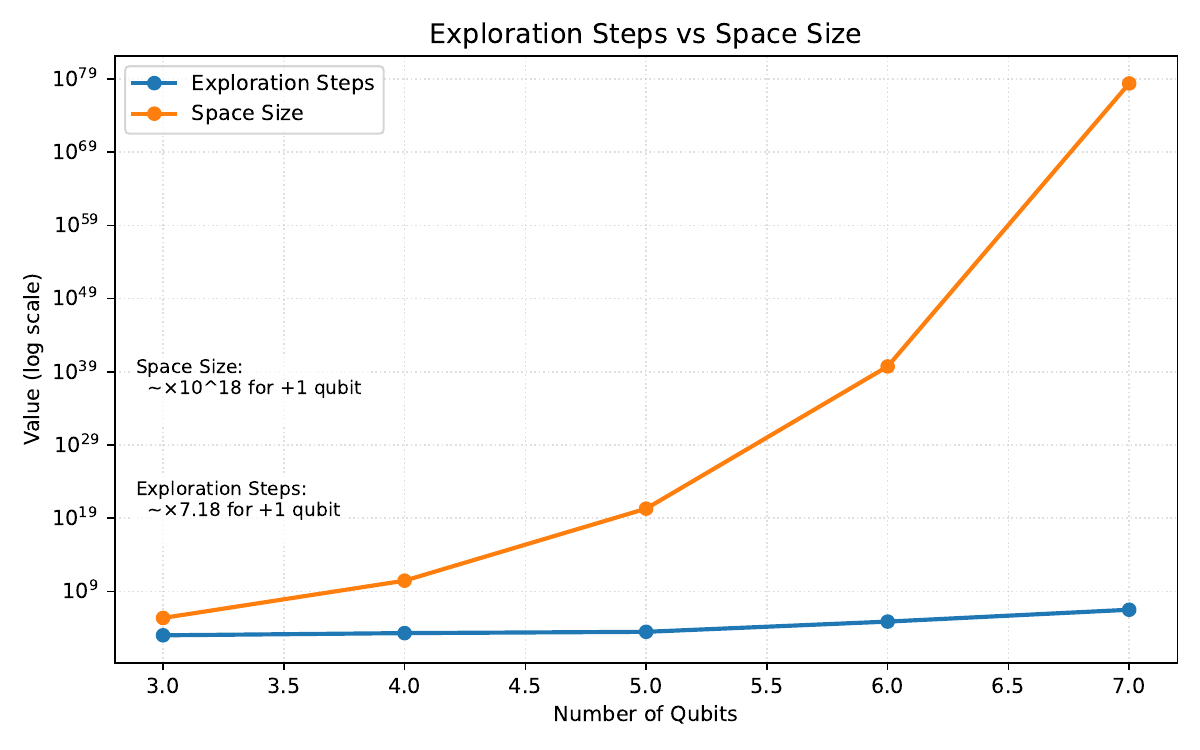}
    \caption{Exploration Steps (blue) and Space Size (orange) as a function of the number of qubits $n$. Both series are shown on the same axes with a logarithmic $y$-scale to make the different scaling regimes directly comparable. The fitted multiplicative growth model $\log_{10} y(n) \approx A + B n$ indicates an average per-qubit growth factor of $\sim 7.2\times$ for the Exploration Steps, but $\sim 3.4\times10^{18}\times$ for the Space Size. While the Exploration Steps increase by about $3.5$ orders of magnitude between $n=3$ and $n=8$, the Space Size explodes by more than $73$ orders of magnitude over the same interval, highlighting the combinatorial blow-up of the search space.}
    \label{fig:Expl_vs_Space}
\end{figure}

To assess how our method scales with the number of qubits $n$, we compare the Exploration Steps and the Space Size in Fig.~\ref{fig:Expl_vs_Space}. Each exploration step corresponds to one Q-matrix update, i.e., the discovery of a new portion of the search environment. Since the number of exploration steps is proportional to the portion of the environment effectively explored, it provides an operational measure of the search effort required to identify the optimal circuit. We observe that this exploration effort grows significantly more slowly than the full state space of SWEET states.  This indicates the validity of our method, even if the exhaustive exploration of the whole state space remains intractable (see Section\ref{subsec:q_learning_for_circuits} for the number of states $N_S$ and actions $N_A$).

\begin{figure}[t]
    \centering
    \includegraphics[width=1\linewidth]{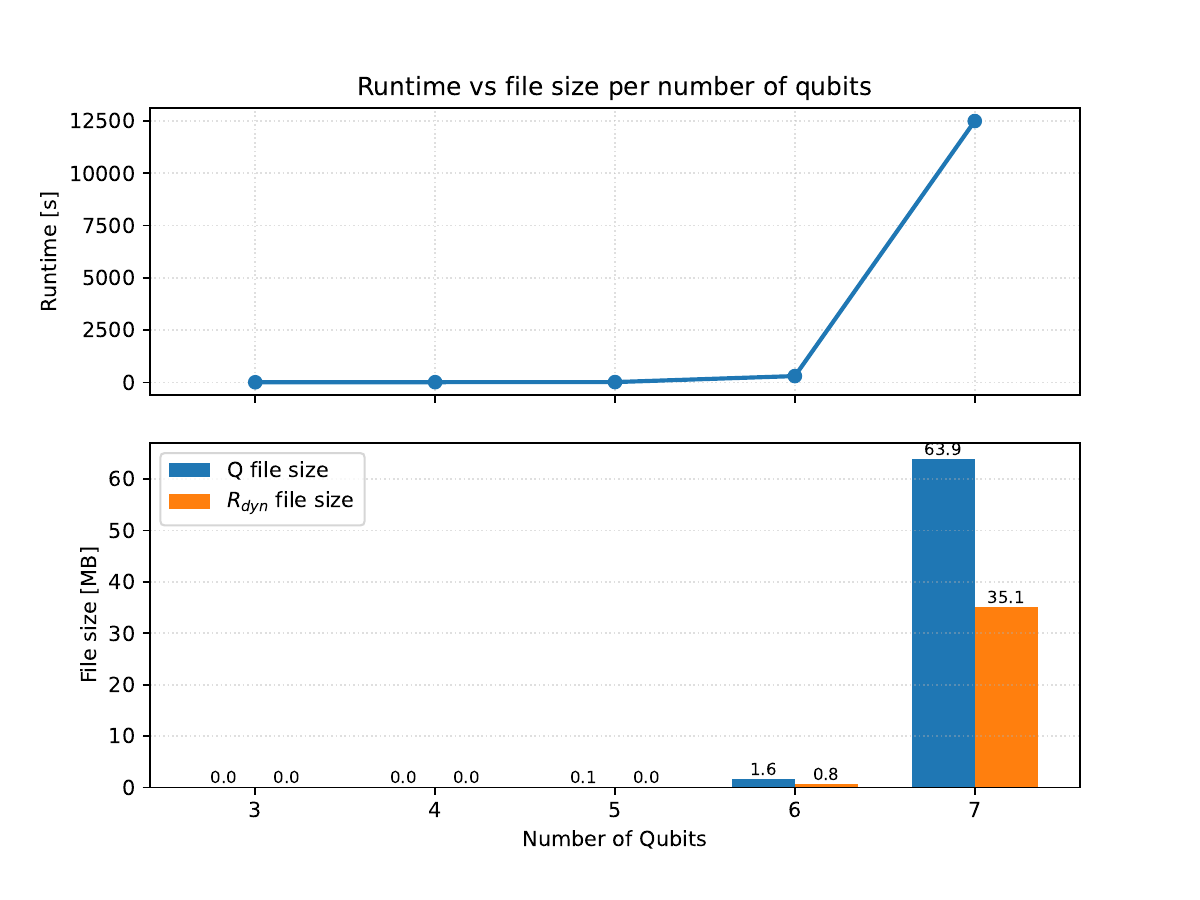}
    \caption{Scaling of computational and memory cost with the number of qubits. (\textbf{Top}) Total runtime required to obtain the optimal circuit with respect to the number of qubits involved in the circuit is shown, where the runtime is computed from the number of exploration steps assuming $4$ ms per step. (\textbf{Bottom}) Storage cost of the learned data structures shown considering the sizes (in MB) of the Q and $R_{\text{dyn}}$ databases produced during training. This joint view shows that, although the runtime grows with the number of qubits, the corresponding storage overhead remains moderate and well below the combinatorial growth of the underlying state space.}
    \label{fig:Memory_and_time}
\end{figure}

For each quantity $y(n)$ we fit a multiplicative growth model of the form $y(n) \approx C \cdot r^{n}$ by performing a linear regression on $\log_{10} y(n)$, i.e.\ $\log_{10} y(n) \approx A + B n$, so that the per-qubit growth factor is $r = 10^{B}$. This procedure yields a characteristic growth factor of approximately $r \approx 7.2$ for the Exploration Steps, meaning that adding one qubit multiplies the required exploration effort by a factor of $\sim 3.5 $ on average. In contrast, for the Space Size we obtain $r \approx 3.4 \times 10^{18}$, i.e.\ each additional qubit increases the size of the search space by roughly $10^{28}$ times. Equivalently, over the interval from $n=3$ to $n=8$ qubits, the Exploration Steps increase by only $\sim 3.5$ orders of magnitude, whereas the Space Size spans more than $73$ orders of magnitude. This dramatic separation in scaling behavior is visually confirmed in Fig.~\ref{fig:Expl_vs_Space}, where both curves are plotted against $n$ with a logarithmic $y$-axis: the Exploration Steps grow moderately, while the Space Size exhibits an explosive, effectively super-exponential expansion. Preliminary results for $n=8$ suggest that the method remains effective and practical due to its reward-guided exploration.

Moreover, we show in Fig.~\ref{fig:Memory_and_time} the memory usage and the running time for growing number of qubits $n$. We are considering the running time only of the training part, not including the one-time offline static reward construction. With our computer architecture each exploration step is $\sim4 \unit{\milli\second}$.

\subsection{General results using universal gate set} \label{subsec:general_result_intro}
This RL technique can be used to create any SWEET state and can be customized to optimize different circuit features. The purpose of this section is indeed not only to show the algorithm’s ability to target states other than graph states, but also to demonstrate its flexibility in incorporating various constraints through adaptable dynamical penalties, depending on the requirements.

In order to generalize the algorithm, our list of actions will consist in gates from the universal gates set $\mathcal{U}=\{H,\text{CNOT},T,T^\dagger\}$ as explained in Section\ref{sec:universal_gates}(\cite{Nielsen_Chuang_2010, Ga_MA_2002}). To implement the operation of $T$ and $T^\dagger$ we need to represent in the SWEET states $M=2^3$ different phases (see Eq.\eqref{eq_SWEET}), thus we include three phase qubits ($p=3$) (\cite{amy2013meet,Glou_2012}). We choose to target a state with $n = 3 $ qubits such that the set of all possible SWEET states has a dimension $ N_S $ on the order of $10^{19}$. The target state is purposely selected to require $T$ or $ T^\dagger $ gates in its generating circuit, starting from the initial state $\ket{\psi^{\text{in}}_3} = \ket{000} $, in order to target a nontrivial circuit. The chosen target state is as follows: 
\begin{equation}\label{eq_target_for_universal_gates}
    \ket{\psi_3}=\ket{010}+\ket{011}+\ket{100}.
\end{equation}
One can notice that both $\ket{\psi_3^{in}}$ and $\ket{\psi_3}$ belong to the SWEET states set. 

\begin{figure}[t]
    \centering
    \includegraphics[width=1\linewidth]{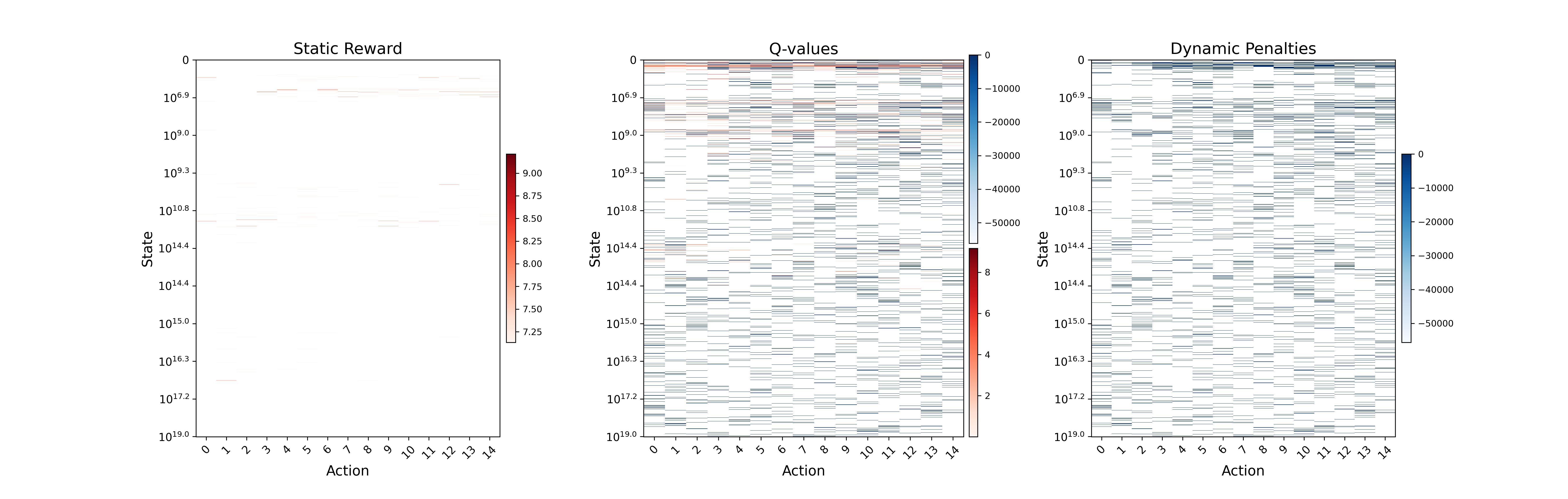}
    \caption{Visualization of the core Q-learning matrices after training on the state $\ket{\psi_3}$ (see Eq.\eqref{eq_target_for_universal_gates}). This figure is analogous to Fig.~\ref{fig:Q_learning_plots_4_qubits}, but here the much larger state-action space makes its structure more apparent. The left panel shows the sparse static reward matrix $R_{\text{sta}}$, which assigns positive ``breadcrumb trail" rewards to actions that lead toward the target state.
    The central panel displays the final Q-matrix after training. Positive values correspond to preferred actions, while negative values indicate actions to be avoided.
    The right panel presents the dynamic penalty matrix $R_\text{dyn}$, which captures the negative rewards used during training to discourage inefficient behavior.
    Together, the static and dynamic reward components shape the Q-matrix, guiding the agent toward an efficient policy. Warm colors denote positive values, and cool colors denote negative values. To enhance the clarity of the visualization, only the non-zero elements of the matrices are depicted using colors. The regions representing the elements of the matrices with zero values are left blank. For further details, refer to Sec.~\ref{subsec:training_phase_universal_gates}. 
    }
    \label{fig:Q_learning_plots_3_qubits}
\end{figure}
\subsection{Designing the reward}

In order to highlight the capability of the algorithm to be customized for different optimization objectives, we first consider generating two circuits: 
\begin{enumerate}
    \item \label{circuit1} a circuit involving the minimum number of gates and
    \item a circuit with optimized depth.
\end{enumerate}

For the first case, we build the static reward matrix with the reward strata discussed in~\ref{subsec:r_static_dynamic}, while, in the dynamic reward, we only include the basic penalties described in Eq.\eqref{eq_reward_penalty}. For the second case, we include the depth penalty along with the basic penalties, with the same method used for graph states (see Eqs. \eqref{eq_congestion_level} and \eqref{eq_depth_penalty}). We consider $R_{\text{max}}=10^4$ and four reward strata, i.e., $k_{\text{max}}=4$.
\subsection{Training phase and sparseness of the matrices}
\label{subsec:training_phase_universal_gates}
We recall that the total number of qubits, including phase qubits, is $n+p=6$ and thus $N_S$ is of the order of $10^{19}$ and the number of actions is $N_A=15$ (see Eqs.\eqref{eq_N_S} and \eqref{eq_N_A}). Despite the vastness of the state-action space, the algorithm is able to construct efficient circuits by exploring only a relevant subspace--sufficient to achieve the objective--while effectively avoiding inconclusive or inefficient paths. This can be seen in Fig.~\ref{fig:Q_learning_plots_3_qubits}, where we depict a projection plot of the R and Q matrices generated at the end of the training phase, for the first type of circuit design, only including basic dynamic penalties. The plot presents a linear scale for the states on the vertical axes, for which the tick labels represent the order of magnitude of the index $s$. On the horizontal axes are represented the actions indexes. This plot is analogous to the plot in Fig.~\ref{fig:Q_learning_plots_4_qubits}, but shows the result of the training phase for a much larger space of $10^{20}$ state-action pairs, showing both zero and nonzero values of the sparse matrices.

The left panel of the plot represents the static reward matrix $R_{\text{sta}}$, which is evidently extremely sparse. It includes only positive values, represented using warm colors. The right panel illustrates the nonzero values of the dynamical reward, $R_{\text{dyn}}$, i.e. the penalties applied dynamically during the training. It only includes negative values, represented using cool colors; the brighter the color, the more penalized the agent would be for choosing the corresponding state-action pair. Although the dynamical rewards are spread all over the space, the matrix is again extremely sparse considering the magnitude of the vertical axis values. The central panel shows the Q-matrix. It consists mostly of negative values along with a few positive ones, constituting the agent policy which guides towards the target state. The static rewards contributed to the positive values of the Q-matrix, whereas the dynamical reward provided negative values, according to Eq.\eqref{eq_q_update}. We recall that, unlike the plot in Fig.~\ref{fig:Q_learning_plots_4_qubits}, this one displays the three matrices on their original scale; therefore, positive values appear in the Q-matrix, in the same region of the R-matrix where $ R_{\text{sta}}(s,a)$ is also positive. The sparsity of the Q-matrix, clearly visible in the plot, indicates that the agent did not explore the entire space during training but rather focused on specific regions, which led to the fast convergence of the algorithm.

\subsection{Optimized circuits with the universal set of gates}\label{subsec:optimized_circuits_universal_set}
\begin{figure}[t]
    \centering
    \includegraphics[width=1.0\linewidth]{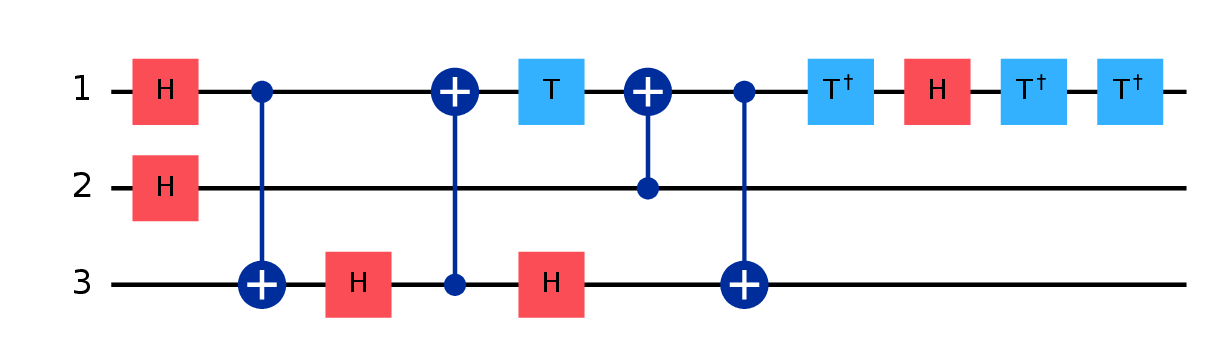}
    \includegraphics[width=0.7\linewidth]{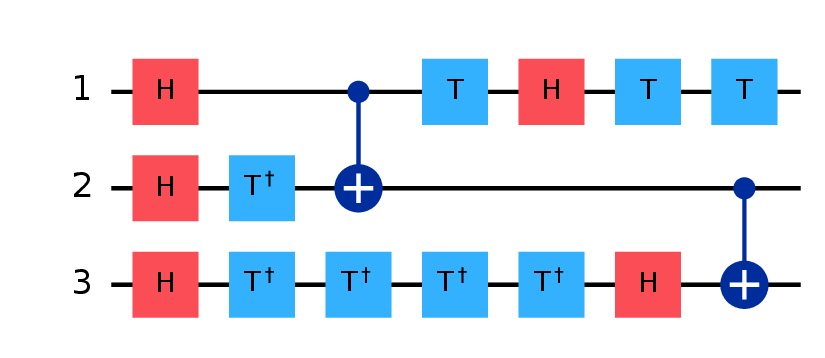}
    \caption{Optimized quantum circuits for preparing three-qubit states. The figure compares two circuit designs, each optimized according to a different criterion. Top Panel: Displays the circuit produced when the optimization prioritizes a minimal gate count. This configuration results in a circuit with 13 gates and a depth of 11. 
    Bottom Panel: Shows the outcome when a depth penalty is included in the optimization, directing the agent to minimize circuit depth. This strategy yields a circuit with a depth of 7  at the expense of a modestly increased gate count (15 gates). These results highlight the effectiveness of dynamic penalty strategies in circuit synthesis, enabling flexible prioritization between gate count and depth according to user-defined objectives. For further methodological details, refer to Sec.~\ref{subsec:training_phase_universal_gates}.
    }
    \label{fig:circuits}
\end{figure}

The designs of the final circuits determined during the testing phase are illustrated in Fig.~\ref{fig:circuits}. The top panel shows the circuit found using the basic penalties defined in Eq.\eqref{eq_reward_penalty}. For this case, we trained the agent over 30$\times 10^3$ episodes of length 50, thus around 15$\times 10^5$ training steps. 
One can notice that the circuit built through the algorithm includes only 13 gates; however, the depth of the circuit is 11. To optimize the depth, we included the depth penalties described in Subsection\ref{subsec:reward_and_pen_for_graph_states}, and constructed the circuit shown in the bottom panel of the same figure. Since the first circuit required only 13 gates, for the latter case, we used episodes of shorter length, that is, of length 30 instead of 50. Formulating an efficient circuit required training over 72$\times 10^3$ episodes, thus 21.6$\times 10^5$ training steps, comparable to the training required with the basic penalties.

As can be noticed from Fig.~\ref{fig:circuits}, the inclusion of the penalty successfully reduced the depth of the circuit from 11 to 7, at the cost of incorporating more gates, which increased from 13 to 15. Hence, these examples further prove the efficiency of the algorithm in modeling and shaping the circuits according to the user's requirements.

To avoid misinterpretation of the results, it is important to clarify the relationship between the final quantum state produced by the optimized circuit and the targeted SWEET state. During training, when the agent encounters non-SWEET states, it maps them onto the discretized SWEET states, an essential step for efficient exploration of the Hilbert space. Such non-SWEET states may arise, for instance, through the application of non-trivial gates such as Hadamard gates. As a result, the final state generated by the circuit may not be exactly equal to the originally targeted SWEET state, but is associated with it via the discretization process. In this sense, the targeted SWEET state serves as the representative of the actual output state produced by the optimized circuit.

The optimal circuits shown in Fig.~\ref{fig:circuits}, which were generated using the Q-learning technique and applied to the predefined initial state $\ket{\psi_3^{\text{in}}}=\ket{000}$, do not produce the expected target state $ \ket{\psi_3} $, but instead produce a different state: $\ket{\psi_3'} = \frac{1}{2}(\ket{010} + \ket{011} + \sqrt{2}\ket{100})$. Nevertheless, since this state is represented and labeled by the targeted SWEET state $ \ket{\psi_3} $, the algorithm is still considered successful (a more detailed discussion is provided in Appendix~\ref{appendix1}). Moreover, one can check the fidelity, $|\bra{\psi'_3}{\psi_3}\rangle|^2$, of the target state, $\ket{\psi_3}$, and the produced state, $\ket{\psi'_3}$, is $0.97$, which is very close to one.

Even if we did not treat this matter more extensively in this work, our intuition is that the non-trivial gate that brings non-SWEETness into the states is the Hadamard gate. Then we could devise methods to increase the fidelity which involve giving penalties for using this specific gate, so that the circuit tries to avoid unnecessary or overuse of Hadamards, which can push the prepared state away from the targeted state. Other future possibilities to overcome this issue are discussed in the conclusions.

 \section{Conclusions}\label{sec:5}
The efficient synthesis of quantum circuits remains a critical challenge to fully harness the potential of quantum computing. This work introduces a hybrid reward-driven reinforcement learning framework that successfully navigates the vast and sparse search space of quantum operations to discover resource-efficient circuits.

The reinforcement learning method involves a hybrid reward system that synergistically combines a precomputed static reward with dynamic penalties generated during the training phase. The static component creates a ``breadcrumb trail" toward the target state due to its layered structure. Concurrently, dynamic penalties provide the flexibility to introduce user-defined constraints, such as minimizing circuit depth or costly T-gate counts, steering the agent toward genuinely efficient solutions.

To avoid wandering through the exponentially large space of state-action pairs, the Q-learning framework includes both exploration and exploitation of the space. To effectively balance between them, we used the well-known $\varepsilon$-greedy method, which efficiently guided the agent to the target regardless of the enormous space dimension. 

To render the problem tractable and usefully discretize the state space, we focused our framework on States With Equal-amplitude and Encoded-phase Terms (SWEET states). In particular, we consider only the SWEET states as the target states for creation of the optimal circuits.  This restriction comes with the limitation of not being able to construct states with unequal and more general amplitudes. However, since the phase can be varied, the SWEET set includes various significantly useful states, such as the maximally entangled states, every graph state, many magic states, etc.
We benchmarked our algorithm by producing circuits for 7-qubit graph states, minimizing circuit depth. The circuits constructed by the algorithm comprise the theoretically predicted minimum depth, according to the Vizing's theorem (\cite{MISRA1992131}). This success demonstrates the algorithm ability to learn and exploit parallelizable gate structures implicitly, from the reward signal alone.

The versatility of the framework is further confirmed by synthesizing SWEET states using a universal gate set ($\{H$, CNOT, $T$, $T^\dagger\}$), proving its adaptability beyond structured problems. The performance of the algorithm is tested for constructing circuits acting on 3-qubit states when the considered number of phase qubits is 3. While for graph states we have always fidelity $1$ between the target state and the one obtained through the optimized circuit, using the universal set of gates the final state reached through the optimized circuit is not exactly the target but is very close to it, with fidelity $0.97$. In our intuition, in the universal set of gates the Hadamard gate is responsible for transforming a SWEET state into a non-SWEET one. Thus, one possibility to construct the exact targeted state or to further increase the fidelity, one can tweak the gates of the optimal circuit appropriately as discussed in our previous work \cite{RL4qubits}. More generally speaking, we believe that the choice of a discretized set of gates brings to this state mismatch. In our future work, we aim to improve and systematically control the fidelity by incorporating continuous parameter optimization, introducing parameterized gates within the already sub-optimal circuit structures generated by our algorithm.

While our present contribution is deliberately tabular, the next step is to lift the same circuit-aware reward design into a deep-RL (DRL) agent. This requires re-engineering the reward distribution so that layered positive signals (progress toward the SWEET class) and strictly negative circuit-quality penalties remain aligned with circuit optimization even when combined with fidelity-based criteria or other mechanisms that tune state amplitudes. In fact, the goal of our work is not to claim scalability, but to provide the reward mechanism and empirical evidence that justify a DRL extension designed around circuit-based objectives rather than fidelity-based optimization. The sparse matrix storage and layered reward shaping make the approach scalable to larger systems while keeping computational overhead manageable. 

The restriction to equal-amplitude components can be mitigated in post-processing by suitably modifying selected gates within the optimized circuit, as proposed in our previous work~\cite{RL4qubits}. A natural direction for future research is to embed this idea within a DRL framework, allowing the agent to manage a wider state space and to design optimized circuits for more general target states beyond the SWEET set.

We stress that DRL methods for state preparation do not map the full Hilbert space; they scale by optimizing over lower–dimensional control spaces (pulse schedules, ansatz parameters, discrete gate sequences) and by learning from compact or proxy measurements, or from structured representations (e.g., tensor networks), while exploring only promising branches of the search space. Seen in this light, our explicit discretization over SWEET states is the tabular analogue of those reductions.

Although the examples presented here involve minimizing the depth of the circuit, the very same penalty-based method can be used to penalize circuits with other faults, such as reducing the number of non-Clifford gates or entangling gates. Importantly, this work highlights how reinforcement learning can integrate classical reward shaping and quantum circuit constraints to navigate the challenges posed by the vastness of state space. 

The proposed algorithm lends itself to a wide range of applications, both within the context of quantum circuit optimization and beyond. One particularly promising direction involves converting it into a circuit synthesizer, targeting unitary transformations rather than a fixed target state. This would leverage the fact that a unitary operator is fully characterized by its action on the computational basis, opening possibilities for discovering minimum-depth decompositions of known quantum gates.

A more immediate extension of our current algorithm concerns the analysis of the penalty patterns that we produce during the training part. Specifically, the $R_{\text{dyn}}$ values constitute penalty maps, which are generated for a given target state. These maps could be generalized to construct informative initial penalty patterns when dealing with larger systems. This approach aims to provide a more scalable variant of the technique by establishing a structured initialization of the agent policy, potentially improving performance for higher qubit counts by accelerating the convergence of the learning procedure.

\bmhead{Acknowledgements}

The authors acknowledge support from Spanish MICIN grant PID2021-122547NB-I00 and the“MADQuantum-CM”project funded by Comunidad de Madrid (Programa de acciones complementarias) and by the Ministry for Digital Transformation and of Civil Service of the Spanish Government through the QUANTUM ENIA project call –Quantum Spain project, and by the European Union through the Recovery, Transformation and Resilience Plan Next Generation EU within the framework of the Digital Spain 2026 Agenda, the CAM Programa TEC-2024/COM-84 QUITEMAD-CM. This work has been financially supported by  the project MADQuantum-CM, funded by Comunidad de Madrid (Programa de Acciones Complementarias) and by the Recovery, Transformation and Resilience Plan—Funded by the EuropeanUnion—(NextGeneration EU, PRTR-C17.I1). KS and SG acknowledge support from the project MADQuantum-CM, funded by Comunidad de Madrid (Programa de Acciones Complementarias) and by the Recovery, transformation and Resilience Plan—Funded by the EuropeanUnion—(NextGeneration EU, PRTR-C17.I1). M.A. M.-D. has been partially supported by theU.S.Army Research Office Through Grant No. W911NF-14-1-0103.

\begin{appendices}

\section{Targeting SWEET states}
\label{appendix1}
As we mentioned in Subsections~\ref{subsec:SWEET} and~\ref{subsec:optimized_circuits_universal_set}, due to the infeasibility of labeling and representing each state of a given Hilbert space in an efficient computational way, we have restricted ourselves to the set of SWEET states. As a result, our algorithm can only register states having the SWEET states form. However, due to the inclusion of non-trivial gates in the algorithm, an action over a SWEET state can bring to a non-SWEET one. For instance, if we apply the Hadamard gate on the first qubit of the SWEET state, $\ket{\psi}=\frac{1}{\sqrt{3}}(\ket{000}+\ket{001}+\ket{010})$, the final state we get is $\ket{\psi'}=\frac{1}{\sqrt{6}}(2\ket{000}+\ket{010}+\ket{011})$, which is not SWEET anymore. 

If we want to build the algorithm for such nontrivial gates, still being restricted to the set of SWEET states, $\sum_j\alpha_j\ket{x_j}$ (see Eq.\eqref{eq_SWEET}), we must consider each SWEET state to represent a class of states, $\sum_ja_j\alpha_j\ket{x_j}$, where $a_j$ are positive real numbers for all $j$s. In these cases, the algorithm treats any state falling outside the SWEET set but within a SWEET class as equivalent to the representative SWEET state of that class. As a result, in such situations, the algorithm does not target a specific state, but rather aims at the class of states represented by a given SWEET state, using the representative SWEET state as the target state, and constructs a circuit that produces a state belonging to that class.

In our intuition, the use of Hadamard gates in the universal set of gates in Sec.~\ref{subsec:general_result_intro}, is the cause of this mismatch. Although targeting the SWEET state \( \ket{\psi_3} \), the algorithm constructs a circuit that does not produce that exact state. Instead, it generates a state that is represented by \( \ket{\psi_3} \) and belongs to the class labeled by it. Notably, despite this mismatch, the process yields an exceptional fidelity of $0.97$, underscoring the effectiveness of the approach.

\end{appendices}
\bibliography{sn_bibliography_new}
\end{document}